\documentclass[%
 reprint,
 amsmath,amssymb,
 aps,
]{revtex4-2}

\usepackage{graphicx}
\usepackage{dcolumn}
\usepackage{bm}

\usepackage{orcidlink}

\hypersetup{
    colorlinks,
    linkcolor={red!50!black},
    citecolor={blue!50!black},
    urlcolor={blue!80!black}
}

\begin{document}

\preprint{APS/123-QED}

\title{Enhancing ultracold atomic batteries using many-body resonances}

\author{Duc Tuan Hoang~\orcidlink{0000-0002-8976-7157}}\email{t.hoang@oist.jp}
\affiliation{Quantum Systems Unit, Okinawa Institute of Science and Technology Graduate University, Onna, Okinawa 904-0495, Japan.}
%
\author{Thomas Busch~\orcidlink{0000-0003-0535-2833}}
\affiliation{Quantum Systems Unit, Okinawa Institute of Science and Technology Graduate University, Onna, Okinawa 904-0495, Japan.}
\author{Thom\'as Fogarty~\orcidlink{0000-0003-4940-5861}}
\affiliation{Quantum Systems Unit, Okinawa Institute of Science and Technology Graduate University, Onna, Okinawa 904-0495, Japan.}



\date{\today}

\begin{abstract}
We study the charging performance of a one-dimensional many-body bosonic quantum battery coupled to a harmonic-oscillator charger. In the weak-coupling regime, we show that the battery--charger dynamics can be accurately described by an effective two-level model, which predicts the resonance condition, optimal charging time, stored work, ergotropy, and charging power. We demonstrate that tuning the charger frequency to the interaction-shifted resonance enables complete energy transfer and maximum extractable work. A many-body charging advantage is observed: increasing the particle number reduces the quantum speed-limit time and enhances the charging power with a characteristic $\sqrt{N_B}$ scaling. We further introduce a decomposition of the switching cost into resonance-shifting and excitation contributions, allowing us to quantify the energetic overhead associated with the charging process. Owing to the small charging cost relative to the available charger energy, stronger battery--charger couplings can be employed to significantly boost the charging power while maintaining a low excitation cost. Our results highlight the role of resonance engineering and many-body effects in designing fast and efficient quantum batteries, and provide a promising route toward their realization in ultracold-atom platforms.

\end{abstract}

\maketitle


\section{Introduction}  
Quantum technologies, based on bona fide quantum resources such as coherence, entanglement, many-body correlations, and statistics, are rapidly advancing with the potential to surpass their classical counterparts in sensing, communication, computation, and energy processing \cite{10.1116/5.0083192, koch2023quantum, l39v-jwwz, Menon_2025,Acin_2018,PRXQuantum.2.040328,ferraro2026opportunities}.  A key challenge in this context is the development of efficient protocols for storing, transferring, and extracting energy at the quantum scale. Within this landscape, quantum batteries (QBs) have emerged as a promising paradigm to address this challenge. They are usually modeled as finite-dimensional 
quantum systems that are charged through a unitary process and later deliver work via controlled extraction protocols \cite{PhysRevE.87.042123, RevModPhys.96.031001, PhysRevB.98.205423, PhysRevA.106.022618, PhysRevA.97.022106, PhysRevResearch.6.013038, PhysRevB.100.115142, PhysRevA.107.042419, PhysRevLett.133.150402, PhysRevA.111.012212,Santos2019,sarkar2025fluctuationenergyextractionquantum}. Beyond their technological relevance, QBs provide a platform for exploring the interplay between quantum thermodynamics, many-body physics, and non-equilibrium dynamics. A central objective in this field is to understand how genuine quantum effects can enhance the power and the maximum extractable work, the so-called ergotropy \cite{kzvn-dj7v, PhysRevLett.118.150601, PhysRevResearch.2.023113, PhysRevLett.125.236402,96cm-ktcb, 10.1116/5.0184903,PhysRevLett.134.180401}.

Over the past decade, many theoretical studies have established that collective charging and correlations can yield superextensive enhancements of the charging power, particularly under global operations that generate multipartite entanglement between battery cells \cite{PhysRevLett.118.150601, PhysRevLett.128.140501,PhysRevLett.111.240401, Binder_2015}. Furthermore, many-body phases can stabilize energy storage, while proximity to quantum criticality can boost work extraction due to enhanced susceptibilities \cite{PhysRevB.100.115142,Barra_2022}. While correlations among battery cells can boost the battery performance, they come with the trade-off that the energy can be locked in correlations between the battery and charger, reducing ergotropy and creating a gap between stored and extractable energy \cite{Allahverdyan_2004, PhysRevLett.122.047702, PhysRevLett.133.150402, PhysRevA.111.012212}. Understanding how to simultaneously maximize charging power, ergotropy, and energy-transfer efficiency therefore remains an open problem.

A wide variety of QB architectures have been proposed and partially realized using finite-dimensional systems, including Dicke-type light--matter models \cite{PhysRevLett.120.117702, PhysRevB.109.235432, PhysRevB.105.115405, PhysRevA.100.043833, PhysRevB.98.205423, PhysRevB.99.205437}, spin and spin-chain batteries \cite{PhysRevA.97.022106, PhysRevResearch.6.013038, PhysRevB.100.115142,Barra_2022, PhysRevApplied.14.024092, PhysRevA.103.033715, PRXQuantum.5.030319,GRAZI2025116383}, ultracold atoms in the lattices \cite{PhysRevA.106.022618} and solid-state platforms \cite{PhysRevLett.131.260401, Hu_2022}. Yet, none have been fully realized currently in experimental settings \cite{RevModPhys.96.031001}. These systems have established many of the fundamental principles underlying quantum-enhanced charging. However, realistic quantum platforms often possess continuous degrees of freedom and an effectively unbounded Hilbert space. In such systems, charging dynamics can involve higher-energy excitations, interaction-induced resonance shifts, and nontrivial many-body effects that have no analogue in finite-dimensional models. Understanding how these features influence energy transfer and work extraction is therefore essential for connecting QB theory with experimentally accessible platforms.

To address these questions, we study the suitability of mixtures of ultracold few-body systems in low dimensions in the continuum as a setting for QBs \cite{GarciaMarch_2014, Sowinski_2019, PhysRevLett.81.1543, PhysRevA.97.023623}, where the interplay between many-body interactions and quantum statistics can lead to many new intriguing phenomena \cite{RevModPhys.80.885,MISTAKIDIS20231, RevModPhys.83.1405,DuongAnh-Tai_2025} which may be advantageous for increasing the charging power. These systems have already been identified as interesting platforms for quantum simulation and technologies \cite{bloch2012quantum, cornish2024quantum}, in particular for the exploration of non-equilibrium thermodynamics of quantum systems \cite{Garcia_2016, 10.21468/SciPostPhys.15.2.048}, as they offer highly-controllable experimental settings \cite{He:10, doi:10.1126/science.1201351, RevModPhys.82.1225}. We will show that many-body interactions effects and the infinite Hilbert space of the continuum system can lead to significant enhancements and interesting features with respect to energy storage and work extraction compared to spin-inspired QBs.

In particular, we propose a one-dimensional, many-body bosonic quantum battery model coupled to a harmonic-oscillator charger. This coupling is realized by a sudden quench of the two-body inter-species contact interaction between the two systems, which can be tuned experimentally using Feshbach resonances \cite{RevModPhys.82.1225,PhysRevLett.81.69}. Our focus is on the interplay between the interaction strength and the charger’s trap frequency, which together determine the interaction-induced resonance condition under which maximum energy transfer and the highest ergotropy can be achieved. In addition to numerically simulating the complex quantum many-body process, we also derive an effective two-level model that provides an approximate description of the battery dynamics in the weak-interaction regime. This approximation gives analytical expressions that predict the resonance conditions, the scaling of the optimal charging power and the Mandelstam--Tamm bound on the quantum speed limit time \cite{Deffner_2017,Mandelstam1945}, which sets the characteristic time for optimal charging in this regime. 
A central outcome of our analysis is a many-body speedup: increasing the particle number shortens the time required to reach the first energy-transfer maximum and enhances the charging power. Moreover, we identify a distinction between the bare resonance and the true interaction-shifted resonance, and show that tuning the charger frequency to the latter enables complete energy transfer and maximal ergotropy. By separating the switching cost into resonance-shifting and excitation contributions, we further show that the charging power can be substantially enhanced while maintaining a small energetic overhead. Finally, we explore the breakdown of the two-level description in the strong-coupling regime, where higher excited states become populated and the charging dynamics acquire a genuinely many-body character.

This manuscript is organized as follows. In Sec.~\ref{sec: model}, we briefly review the concept of a quantum battery and the charger-mediated charging protocol. We also introduce the key quantities of interest, including the total stored work, the irreversible work, the maximal extractable work (ergotropy), and the charging power. We then present our quantum battery model using two-component ultracold atoms. Our main results are reported in Sec.~\ref{sec: results}, including the effective two-level model in the weak-coupling regime, the scaling of charging power and the switching cost with increasing particle number, how to optimize the charging power under a charging-cost constraint and the breakdown of the two-level model in stronger interaction limit. 
Finally, our conclusions and outlook are presented in Sec.~\ref{sec: conclusions}. Altogether, our work establishes resonance engineering in interacting continuum systems as a powerful design principle for quantum energy storage, illustrating how quantum statistics, many-body effects, and finite charging costs can be harnessed to optimize the performance of future QBs.

\section{\label{sec: model}Model} 
\subsection{\label{subsec: charger-mediated}Charger-mediated protocol}
We consider a quantum battery, described by the Hamiltonian $H^B$, which is charged through unitary dynamics via the coupling with another quantum system, denoted as a quantum charger $H^C$.  Initially, the Hamiltonian of the uncoupled battery-charger system is given by
\begin{equation} \label{charger-mediated}
    H_0^{BC} = H^{B} + H^{C},
\end{equation}
and the battery is prepared in its ground state $\rho^B_0$, while the charger is in an excited state 
\cite{PhysRevB.98.205423, RevModPhys.96.031001}. The corresponding separable state is given by
\begin{equation}
    \rho^{BC}(t=0) = \rho^B(0)  \otimes\rho^C(0) = \rho^{B}_0\otimes\rho^C(0).
\end{equation}
At $t = 0$, the interaction $H^{\text{int}}$ between the battery and the charger is activated, and the charging process starts. The time-dependent Hamiltonian of the system then becomes 
\begin{equation} \label{charging Hamiltonian}
    H_1^{BC}(t) = H^{BC}_0  + \lambda(t)H^{\text{int}},
\end{equation}
where $\lambda(t)$ is an external classical control that modulates the coupling between the battery and the charger, allowing the composite BC system to exchange energy. 

In our work, charging dynamics are described by a sudden quench of the coupling, 
which are modeled by defining $\lambda(t)$ as a step function 
\begin{equation}
    \lambda(t)=
    \begin{cases}
      1, & \text{if}\ t \in [0,\tau], \\
      0, & \text{otherwise}.
    \end{cases}
\end{equation}
During the charging the state of the charged battery becomes time-dependent and is governed by the Liouville-von Neumann equation
\begin{equation} \label{charged state}
    \dfrac{d}{dt}\rho^{BC}(t) = -\dfrac{i}{\hbar} \left[H_1^{BC},\rho^{BC}(t) \right].
\end{equation} 
At the end of the charging protocol, the interaction is switched off (modeled as another step function) 
and the quantum state of the battery can be obtained by tracing out the  contribution of the charger from the composite density matrix 
\begin{equation}
    \rho^{B}(t) = \text{Tr}_{C}[\rho^{BC}(t)].
\end{equation}

\subsection{\label{subsec: ergotropy}Work extraction process}
After the charging time $t$, the total work stored in the quantum battery is given by
\begin{align} \label{total work eq}
    W_B(t) &= \text{Tr}\{H^{B} \left[\rho^{B}(t) -  \rho^{B}_0\right] \}.
\end{align}
To quantify the maximal extractable work from a battery through a reversible cyclic unitary process, one can compute the ergotropy by optimizing over all possible unitaries $\mathcal{U}$ that lower the energy of the charged state $\rho^{B}(t) = \sum_i \lambda_i |\varphi_i \rangle \langle \varphi_i|$ (with $\lambda_i>\lambda_{i+1}$) with respect to $H^{B} = \sum_i \varepsilon_i |\psi_i \rangle \langle \psi_i|$ (with $\varepsilon_i<\varepsilon_{i+1}$) \cite{Allahverdyan_2004}
\begin{equation} \label{ergotropy-unitary}
    \mathcal{E}_B = \text{Tr}\left[H^{B} \rho^{B}(t)\right] - \text{min}_\mathcal{U} \left\{ \text{Tr}\left[H^{B} \mathcal{U} \rho^{B}(t)\mathcal{U}^{\dagger}\right] \right\},
\end{equation}
which can also be expressed as
\begin{equation}    
    \mathcal{E}_B = \text{Tr}\left[H^{B} \rho^{B}(t)\right] - \text{Tr}\left[H^{B} P_\rho\right].
\end{equation}
In other words, the ergotropy represents the difference between the mean energy
of the state $\rho^{B}(t)$ and the passive energy, defined as the energy of a passive state $P_\rho$, from which no additional work can be extracted via unitary operations. A quantum state is classified as passive when any unitary transformation applied to it increases its energy
\begin{equation}
    \text{Tr}\left[H^{B} P_\rho\right] \leq \text{Tr}\left[H^{B}\mathcal{U} P_\rho \mathcal{U}^{\dagger}\right].
\end{equation}

After the charger is decoupled, the entropy $S(\rho^{B}) = - \sum_i\lambda_i\log{\lambda_i}$ is conserved, and therefore the eigenvalues $\lambda_i$ remain constant. A passive state, which is stationary due to its commutation with $H_0^{B}$, is defined in the diagonal basis of the local Hamiltonian with reordered occupations such that higher occupation numbers correspond to lower energy levels
\begin{equation} \label{passive-state}
    P_\rho = \sum_i \lambda_i |\psi_i \rangle \langle \psi_i|.
\end{equation}
The ergotropy can then be expressed as
\begin{equation} \label{ergotropy eq}
    \mathcal{E}_B = \sum_i (p_i - \lambda_i)\varepsilon_i,
\end{equation}
where $p_i = \sum_j \lambda_j |\langle \varphi_j|\psi_i \rangle|^2,$ is the projection of $\rho^{B}(t)$ onto $H_0^{B}$, representing the energy probability distribution of the charged battery state. For non-zero ergotropy, it is required that $p_i \neq \lambda_i$ \cite{PhysRevResearch.6.013038}. 

In general, the ergotropy $\mathcal{E}_B$ is less than or equal to the total work stored in the battery $W_B$, expressed as $\mathcal{E}_B \leq W_B$. The equality holds when the state $\rho^{B}$ is pure. For a mixed state $\rho^{B}$, a nontrivial gap between $\mathcal{E}_B$ and $W_B$ arises as some energy is locked within correlations between the charger and the battery \cite{PhysRevLett.122.047702, PhysRevB.100.115142, RevModPhys.96.031001}. 
To compare the performance between the many-body battery and its single-particle counterpart, we define the charging power
\begin{equation}
    \mathcal{P}_W = \dfrac{W_B(t_{\text{opt}})}{t_{\text{opt}}},
\end{equation}
where $t_{\text{opt}}$ denotes the shortest (optimal) time for which the maximum work $W_B$ is stored in the battery.

\subsection{\label{subsec: ultracold battery}Many-body ultracold atoms battery}
We will study a charger-mediated protocol by modeling a quantum battery using a two bosonic systems confined in different one-dimensional harmonic traps, where one component acts as the battery, with trap frequency $\omega_B$, and the other as the charger, with trap frequency $\omega_C$. The initial Hamiltonian is expressed as \eqref{charger-mediated}, where
\begin{align}
    H^{B} = \sum_{i=1}^{N_{B}} \Big(&-\dfrac{\hbar^2}{2m} \dfrac{\partial^2}{\partial x_{B,i}^2} + \dfrac{1}{2}m\omega_B^2 x_{B,i}^2 \Big),  \\ 
    H^{C} = -\dfrac{\hbar^2}{2m} &\dfrac{\partial^2}{\partial x_C^2} + \dfrac{1}{2}m\omega_C^2 x_C^2 .
\end{align}
Here, the many-body battery consists of $N_B$ non-interacting bosonic particles and is charged by a single-particle charger, with equal masses $m_B = m_C = m$. 

Without loss of generality we assume that the battery is initially prepared in the ground state of the Hamiltonian $H^B$ with energy $E_B(0)=\frac{N}{2}\hbar\omega_B$ while the charger is prepared in the 1st excited state of the Hamiltonian $H_C$ with energy $E_C(0)=\frac{3}{2}\hbar\omega_C$. To charge the battery, an inter-species interaction term $H^{\text{int}}$ is introduced, which couples the battery and charger 
\begin{equation}
    H^\text{int} = g_{BC} \sum_{i=1}^{N_{B}} \delta(x_{B,i} - x_{C}), 
\end{equation}
and where $g_{BC}$ denotes the short-range inter-species coupling strength. This interaction originates from low-energy $s$-wave scattering between bosons confined in one dimension and is well described by a delta-function potential. In ultracold-atom realizations, the effective interaction strength $g_{BC}$ can be tuned experimentally by adjusting the transverse confinement \cite{PhysRevLett.81.938} or via Feshbach resonances \cite{RevModPhys.82.1225, PhysRevLett.81.69}, providing direct control over the many-body correlations between the battery and the charger. By switching on this coupling, energy is transferred from the charger to the battery, enabling the charging process. 

\begin{figure} [t!]
\center
\includegraphics[width=\columnwidth]{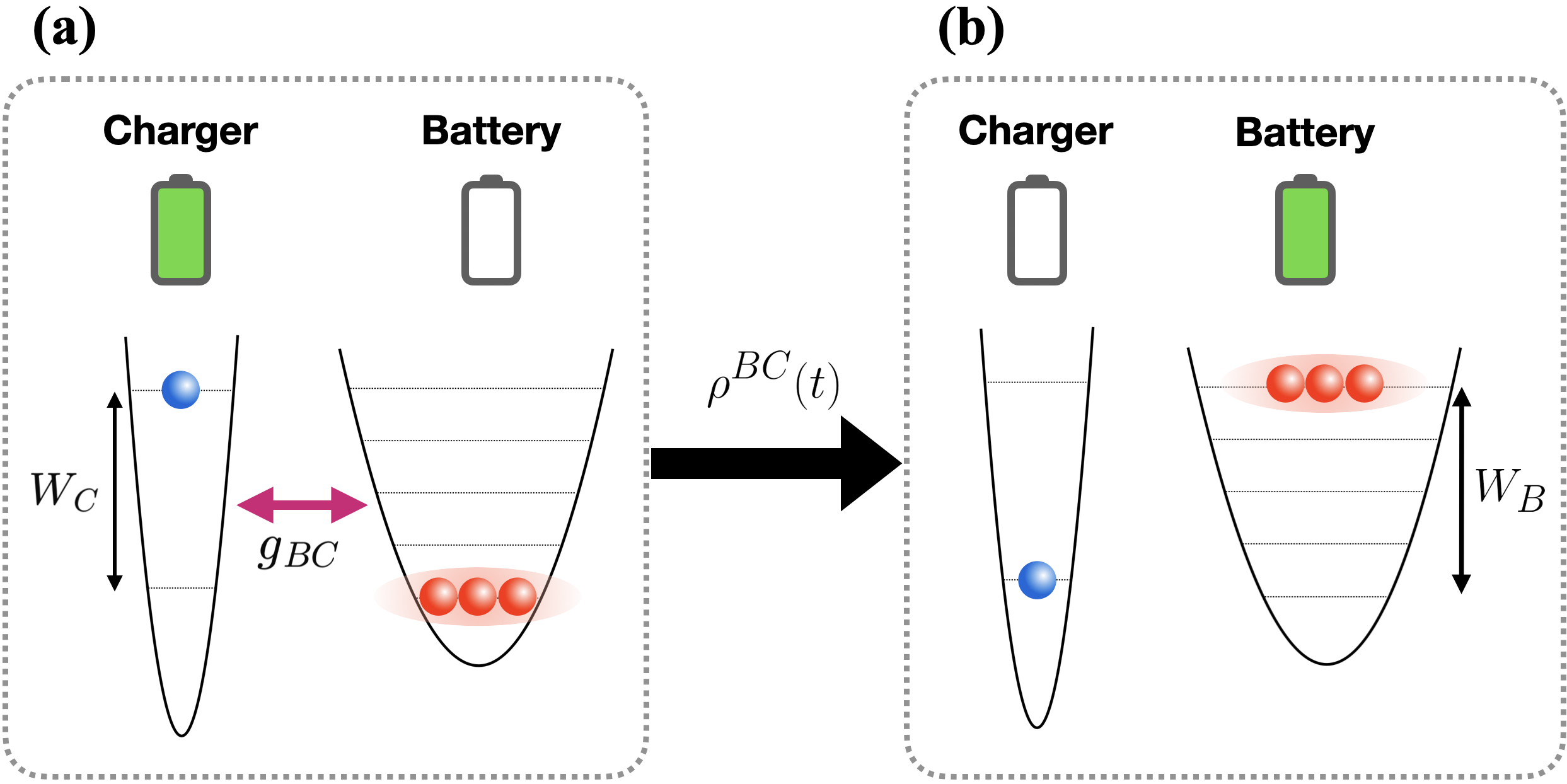} 
\caption{Charger-mediated protocol using a two-component bosonic system in 1D harmonic traps. (a) Initial state with the charger in the first excited state with stored work $W_C$, and $N_B$ bosons in the ground state of the battery. After the coupling is applied with strength $g_{BC}$ the combined system $\rho_{BC}(t)$ evolves to (b) where the charger particle occupies its groundstate while the battery is excited with stored work $W_B$.}
\label{fig: model}
\end{figure}

For the charger, the initial stored work can be varied through the trap frequency $\omega_C$  
\begin{equation} \label{eq: W_C}
   W_C(0) = \varepsilon_1^C-\varepsilon_0^C = \hbar\omega_C,
\end{equation}
where $\varepsilon_i^C$ are the eigenenergies of $H_C$. The work that is stored in the battery is given by
\begin{equation}
    W_B(t) = E^B(t)-\varepsilon^B_0,
\end{equation}
with $E^B(t)=\text{Tr}\left[H^B \rho^B(t)\right]$ the energy of the battery state during the charging dynamics and $\varepsilon_i^B$ are the eigenenergies of $H_B$. 
The model is schematically illustrated in Fig.~\ref{fig: model}.

\section{Results} \label{sec: results}

\subsection{\label{subsec:2-level model}Effective two-level model}
Let us consider a quantum battery system with $N_B$ bosons, which is charged through a weak battery-charger coupling $g_{BC}\ll1$. 
Since the battery consists of identical bosons, the particles are indistinguishable and the excited many-body state must be properly symmetrized. We therefore define the fully charged battery as the bosonic state in which a single excitation to the $n$th excited level is coherently shared among all $N_B$ particles, while the remaining population resides in the ground state. At the same time, the charger is fully discharged, with the corresponding excitation relaxed to its ground state.
In this case we can approximate the system by a two-level model 
\begin{align}
    |0\rangle &= \left[\prod_{i=1}^{N_B} \phi_0(x_{B,i})\right]\varphi_1(x_C), \\
    |1\rangle &= \frac{1}{\sqrt{N_B}}\, \mathcal{S}\!\left[ \phi_{n}(x_{B,1})\phi_{0}(x_{B,2})\cdots\phi_{0}(x_{B,N}) \right]\varphi_{0}(x_C) \label{eq: 1st excited state formula} ,
\end{align}
where $\phi_n(x_B)$ and $\varphi_n(x_C)$ are single-particle wave functions with respect to $H^B$ (with $N_B=1$) and $H^C$,  
respectively, and $\mathcal{S}$ is the symmetrization operator.
Using the analytical expressions of harmonic oscillator eigenfunctions, we can write the matrix form of the charging Hamiltonian, with the four matrix elements are defined as
\begin{align}
   \langle0|H_1^{BC}|0\rangle &= \dfrac{N_B}{2}\hbar\omega_B+\dfrac{3}{2}\hbar\omega_C +N_Bg_{BC}I_{01},\nonumber\\
   \langle0|H_1^{BC}|1\rangle &= \langle1|H_1^{BC}|0\rangle = g_{BC}\sqrt{N_B}\mathcal{I}_n, \nonumber\\
   \langle1|H_1^{BC}|1\rangle &= \left(\dfrac{N_B} {2}+n\right)\hbar\omega_B + \dfrac{1}{2}\hbar\omega_C \nonumber\\
    & \hspace{40pt}+g_{BC}\left[(N_B-1)I_{00} + I_{n0} \right],
    \label{Eq:eff_ham}
\end{align}
with the corresponding integrals
\begin{align}
    I_{00} &= \int \phi^2_0(x)\varphi_0^2(x)dx = \sqrt{\dfrac{m}{\pi\hbar}}\sqrt{\dfrac{\omega_B\omega_C}{\omega_B+\omega_C}},\\
    I_{01} &= \int \phi^2_0(x)\varphi_1^2(x)dx = \sqrt{\dfrac{m}{\pi\hbar}}\dfrac{\omega_C\sqrt{\omega_B\omega_C}}{(\omega_B+\omega_C)^{3/2}},\\
    \begin{split}
        I_{n0} &= \int \phi^2_n(x)\varphi_0^2(x)dx =\sqrt{\frac{m\omega_C}{\pi\hbar}} \left( \frac{\omega_B}{\omega_B + \omega_C} \right)^{n+1/2} \\
    &\quad \times\left( 1 + \sum_{\substack{k=2 \\ k \text{ even}}}^{n} \binom{n}{k} \frac{(k-1)!!}{k!!} \left(\frac{\omega_C}{\omega_B}\right)^k \right),
    \end{split}
     \\
    \mathcal{I}_n &=  \int \phi^*_0(x)\varphi_1^*(x)\phi_n(x)\varphi_0(x)dx. \label{eq: overlap}
\end{align}
The diagonal terms of the effective Hamiltonian quantify the energies of the initial and final states of the total battery-charger system, $|0\rangle$ and $|1\rangle$ respectively. They have two main contributions: the bare energies in terms of the harmonic oscillator frequencies $\omega_B$ and $\omega_C$, and the interaction induced energy shifts which depend on the battery charger coupling $g_{BC}$. The off-diagonal terms describe the coupling between the two states, which drives the energy transfer process. 

The effective two-level charging Hamiltonian can be written in terms of Pauli matrices as 
\begin{equation}
     H_{1,\text{eff}}^{BC} = \dfrac{\delta}{2}\sigma_z+J\sigma_x+\text{const},
\end{equation}
where the detuning $\delta =  \langle0|H_1^{BC}|0\rangle -  \langle1|H_1^{BC}|1\rangle$ is the energy difference between the two diagonal elements, and 
the off-diagonal elements $J = g_{BC}\sqrt{N_B}\mathcal{I}_n$ are the effective coupling terms.
The time-evolved state can be derived as 
\begin{align} \label{eq: psi_t}
    |\Psi(t)\rangle &= \left[ \cos{\left(\dfrac{\Omega t}{2\hbar}\right)} -i\dfrac{\delta}{\Omega}\sin{\left(\dfrac{\Omega t}{2\hbar}\right)}  \right]|0\rangle \nonumber\\
    & \hspace{80pt} -i\dfrac{2J}{\Omega}\sin{\left(\dfrac{\Omega t}{2\hbar}\right)}|1\rangle,
\end{align}
where $\Omega = \sqrt{(2J)^2+\delta^2}$ is the frequency of the charging dynamics. The total amount of work stored in the battery is then expressed as
\begin{equation} \label{eq: W_B}
    W_B(t) =W_C(0)\left(\dfrac{2J}{\Omega}\right)^2 \sin^2\left(\dfrac{\Omega t}{2\hbar} \right),
\end{equation}
while the ergotropy is
\begin{align} \label{eq: ergotropy_B}
    \mathcal{E}_B(t) =\max\Big\{[W_C(0)-\hbar\omega_B]\left(\dfrac{2J}{\Omega}\right)^2\sin^2\left(\dfrac{\Omega t}{2\hbar} \right), \nonumber \\
    [W_C(0)+\hbar\omega_B]\left(\dfrac{2J}{\Omega}\right)^2\sin^2\left(\dfrac{\Omega t}{2\hbar} \right) - \hbar\omega_B \Big\}.
\end{align}

\begin{figure}
\center
\includegraphics[width=0.9\columnwidth]{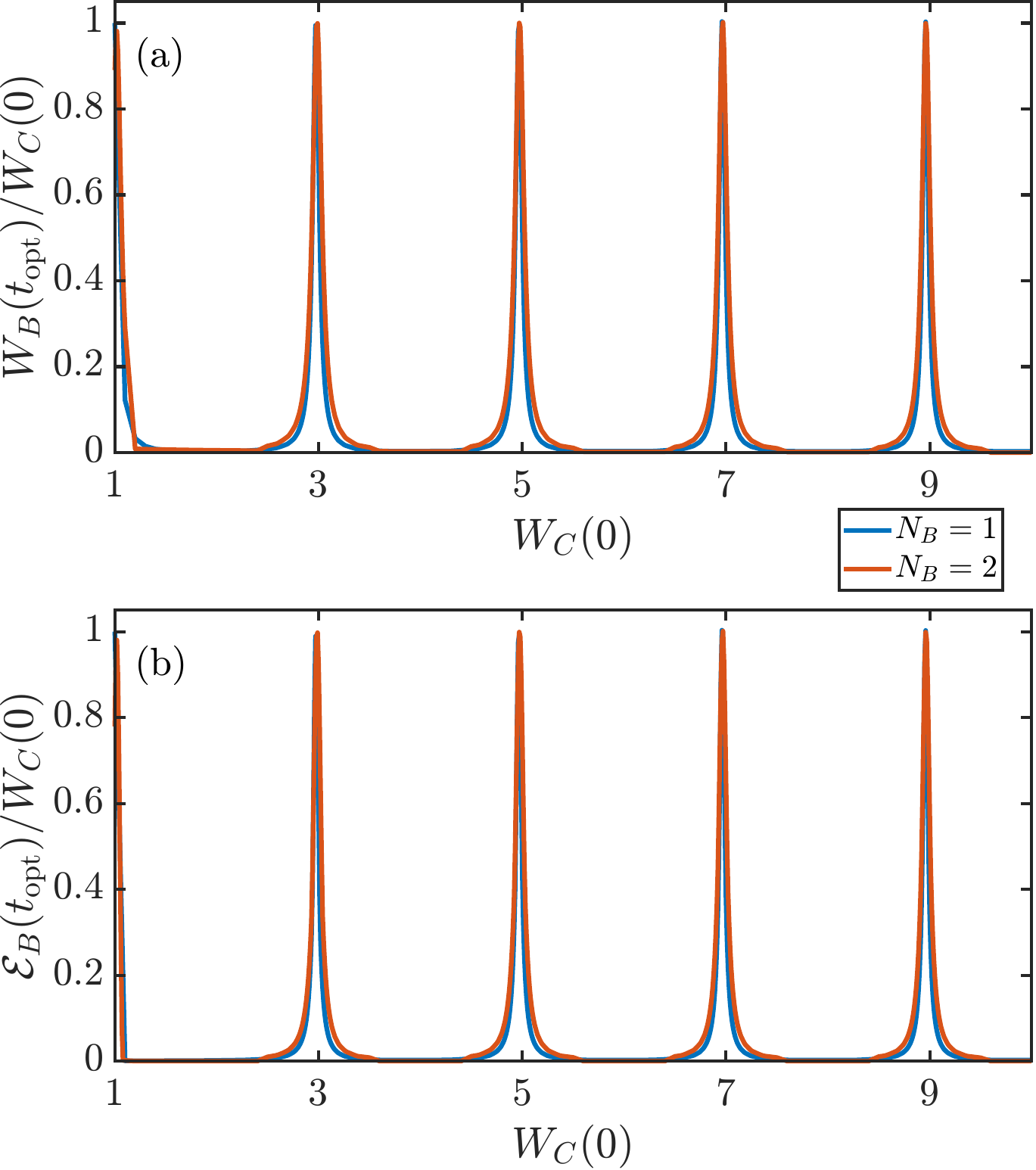} 
\caption{(a) The maximum stored work $W_B(t_{\mathrm{opt}})$ and (b) the corresponding ergotropy $\mathcal{E}_B(t_\mathrm{opt})$ of the battery, both rescaled by the initial stored work of the charger $W_C(0)$. The blue curve represents the single-particle battery, while the red curve corresponds to $N_B = 2$. A coupling is set to $g_{BC} = 0.1$. All numerical data are obtained using the exact diagonalization method (see Appendix~\ref{subsec: ED}).}
\label{fig: resonance spectrum}
\end{figure}

One can immediately see from Eq.~\eqref{eq: W_B} that the optimal charging time to achieve the first maximum energy transfer is
\begin{equation} \label{eq: optimum time}
    t_\text{opt} = \dfrac{\pi\hbar}{\sqrt{(2J)^2+\delta^2}},
\end{equation}
leading to the corresponding charging power 
\begin{equation}
    \mathcal{P}(t_\text{opt})=\dfrac{W_C(0)}{\pi\hbar}\dfrac{4J^2}{\sqrt{(2J)^2+\delta^2}}.
\end{equation}
In our model, we fix the trap frequency of the battery and express all quantities in natural units, setting $\hbar = m = \omega_B = 1$. Figures~\ref{fig: resonance spectrum}(a) and (b) display the maximum stored work and the corresponding ergotropy as functions of the charger's initial energy, $W_C(0)=\omega_C$. For each value of $\omega_C$, we determine the time at which the battery reaches its first maximum in stored energy, denoted by $t_{\text{opt}}$, and compute the corresponding ratios $W_B(t_{\text{opt}})/W_C(0)$ and $\mathcal{E}_B(t_{\text{opt}})/W_C(0)$. Near odd values of $W_C(0)$, both ratios increase and eventually reach unity, indicating perfect energy transfer from the charger to the battery, before rapidly decreasing back to zero.

In contrast, no energy transfer occurs near even values of $W_C(0)$ in the weak-coupling regime, despite the large initial energy available in the charger. This behavior can be understood from the effective coupling term $J$ which depends on the overlap defined in Eq.~\eqref{eq: overlap}. The overlap involves four single-particle states: the initial states consist of the battery ground state $\phi_0(x)$, which has even parity, and the first excited state of the charger $\varphi_1(x)$, which has odd parity; the final states are the charger ground state $\varphi_0(x)$, with even parity, and the $n$th excited state of the battery $\phi_n(x)$. For energy transfer from the charger to the battery to occur, the overlap integral must be nonzero. This requires the product of the four wave functions to be an overall even function. Since the product $\phi_0(x)\varphi_1(x)\varphi_0(x)$ has odd parity, the battery excited state $\phi_n(x)$ must also be odd-parity in order to make the entire integrand even. Consequently, only odd-parity excited states of the battery can participate in the energy-transfer process in the weak-$g_{BC}$ regime.

To gain deeper insight into the charging dynamics, we first consider the case in which the charger frequency is fixed at $\omega_C=n$. For the example discussed below, we choose $n=3$, corresponding to the energy required to excite a particle in the battery to the third excited state. We refer to this condition as the \emph{bare resonance} because, in the absence of battery--charger coupling ($g_{BC}=0$), the charger energy exactly matches the excitation energy of the battery, as seen in Eq.~\eqref{Eq:eff_ham}.
\begin{figure*}
\center
\includegraphics[width=2\columnwidth]{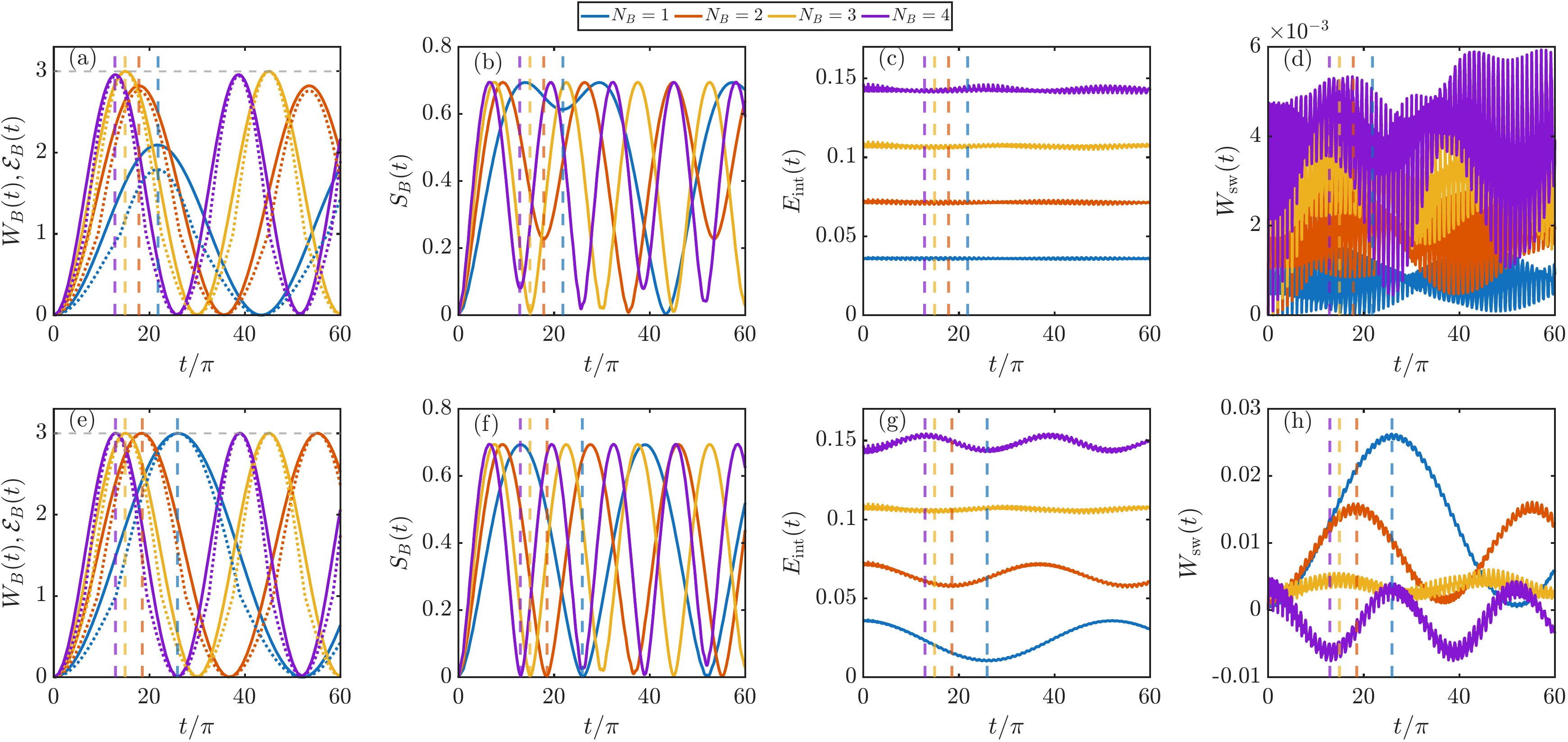} 
\caption{(a),(e) The total stored work in the battery $W_B$ (solid lines) and the ergotropy $\mathcal{E}_B$ (dotted lines); (b),(f) the von Neumann entropy $S_B$ of the battery state $\rho^B(t)$; (c),(g) the interaction energy $E_{\text{int}}$; and (d),(h) the switching cost $W_{\text{sw}}$, as functions of the charging time $t$, for $n=3$ and $g_{BC}=0.1$. The vertical dashed color lines indicate the time at which the energy transferred from the charger to the battery reaches its first maximum while the horizontal gray dashed line represents the maximum work transfer to the battery. Panels (a)--(d) are obtained for the bare resonance $\omega_C=3$, whereas panels (e)--(h) use the resonance frequency $\omega_C^\mathrm{res}$ determined from Eq.~\eqref{eq: delta}. In all panels, the blue curves correspond to the single-particle battery, while the red, yellow and purple curves represent non-interacting two-, three- and four-particle batteries, respectively.}
\label{fig: dynamics}
\end{figure*}

The resulting charging dynamics are shown in Fig.~\ref{fig: dynamics}(a)--(d). Figure~\ref{fig: dynamics}(a) displays the total stored work $W_B$ and the corresponding ergotropy $\mathcal{E}_B$ for batteries containing up to $N_B=4$ particles. For clarity, it is useful to distinguish three regimes: $N_B<n$, $N_B=n$, and $N_B>n$. For $N_B<n$, only a fraction of the charger's initial energy is transferred to the battery. In particular, for $N_B=1$, approximately two-thirds of the available energy is stored at the optimal charging time $t_{\text{opt}}$. Moreover, a noticeable gap remains between the stored work and the ergotropy, indicating that the battery state remains significantly mixed due to correlations with the charger. Increasing the number of particles in the battery not only enhances the amount of energy transferred from the charger to the battery, but also reduces the gap between the stored work and the ergotropy. This trend continues until $N_B=n=3$, where the gap essentially vanishes and the transferred energy reaches its maximum value. However, for $N_B>3$, the transferred energy begins to decrease again and the gap between $W_B$ and $\mathcal{E}_B$ widens. It is worth noting that increasing the particle number generally shortens the optimal charging time, irrespective of the amount of transferred energy.

The reduction of the ergotropy gap as $N_B$ approaches $n$ can be understood by examining the time-dependent von Neumann entropy of the battery state $\rho^B(t)$, shown in Fig.~\ref{fig: dynamics}(b). At the optimal charging time corresponding to the first maximum in energy transfer, the entropy remains finite for $N_B<n$, indicating that the battery state is mixed. Smaller particle numbers result in larger entropy at $t_{\text{opt}}$, implying stronger correlations between the battery and the charger, which reduce the extractable work and degrade the battery performance. Remarkably, for $N_B=n=3$, the entropy becomes essentially zero at $t_{\text{opt}}$, indicating that the battery state is pure and uncorrelated with the charger. Consequently, the stored work and the ergotropy become nearly identical. For $N_B>3$, correlations between the battery and the charger build up more rapidly, leading to faster energy transfer. However, the entropy again remains finite at $t_{\text{opt}}$.

The reduced charging time for larger batteries can also be understood by examining the interaction energy,
$
E_{\text{int}} = \text{Tr}[H^{\text{int}}\rho^{BC}(t)],
$
shown in Fig.~\ref{fig: dynamics}(c). For a fixed coupling strength, increasing the number of particles increases the interaction energy within the battery--charger system. These collective interactions, due to the charger coupling to all particles in the battery simultaneously, result in reducing the characteristic charging timescale even though the interaction strength is fixed. The high-frequency oscillations visible in the interaction energy are associated with a breathing mode induced by the repulsive battery--charger interaction, with a characteristic period on the order of $\pi/\omega_B$. These oscillations lead to small density modulations in both the battery and charger states.

Notably, since the interaction Hamiltonian $H^{\text{int}}$ does not commute with the bare Hamiltonian $H_0^{BC}$, i.e.
$
[H_0^{BC}, H^{\text{int}}] \neq 0,
$
switching on the interaction contributes additional energy to the system during the charging process. Consequently, the energy stored in the battery is not transferred entirely from the charger, but may also contain a contribution from the switching cost,
\begin{equation} \label{eq: Wsw}
     W_{\text{sw}} = \text{Tr}\{H^\text{int}\left[\rho^{BC}(0)-\rho^{BC}(t)\right]\}\,,
\end{equation}
which is defined as the energy difference between switching on and off the interaction term at the beginning and the end of the charging process~\cite{PhysRevB.98.205423}. This quantity is shown in Fig.~\ref{fig: dynamics}(d). We find that the switching cost remains extremely small compared to the initial energy stored in the charger, contributing less than $0.1\%$ throughout the charging protocol. Therefore, the stored work in the battery originates almost entirely from energy transferred by the charger, while only a negligible fraction is associated with switching-induced excitations and irreversible work.

The observation that perfect energy transfer is achieved only for $N_B=n$ suggests that the bare resonance does not generally coincide with the true resonance of the interacting system. Indeed, once the battery and charger are coupled, the interaction modifies the energies of both the initial and charged states, shifting the resonance away from the bare value $\omega_C=n$.
The interaction-corrected resonance can be obtained from the two-level model by introducing the detuning
\begin{equation} \label{eq: delta}
\begin{split}
\delta &= \omega_C - n + \frac{g_{BC}}{(1+\omega_C)^{n+1/2}}\sqrt{\frac{\omega_C}{\pi}} \\
&\quad \times \left[P_n(\omega_C)- N_B(1+\omega_C)^{n-1} \right],
\end{split}
\end{equation}
where
\begin{equation}
    P_n(\omega_C)=(1+\omega_C)^n-1-\sum_{\substack{k=2 \\ k \text{ even}}}^{n} \binom{n}{k} \frac{(k-1)!!}{k!!} \omega_C^k.
\end{equation}
The \emph{true resonance} frequency $\omega_C^{\mathrm{res}}$ is obtained by solving for the condition $\delta = 0$. 
At true resonance, the generalized Rabi frequency satisfies $\Omega = 2J$, which results in complete energy transfer from the charger to the battery. Consequently, both the stored work and the ergotropy attain their maximum values, as shown by Eqs.~\eqref{eq: W_B} and \eqref{eq: ergotropy_B}.

\begin{figure}
\center
\includegraphics[width=0.8\columnwidth]{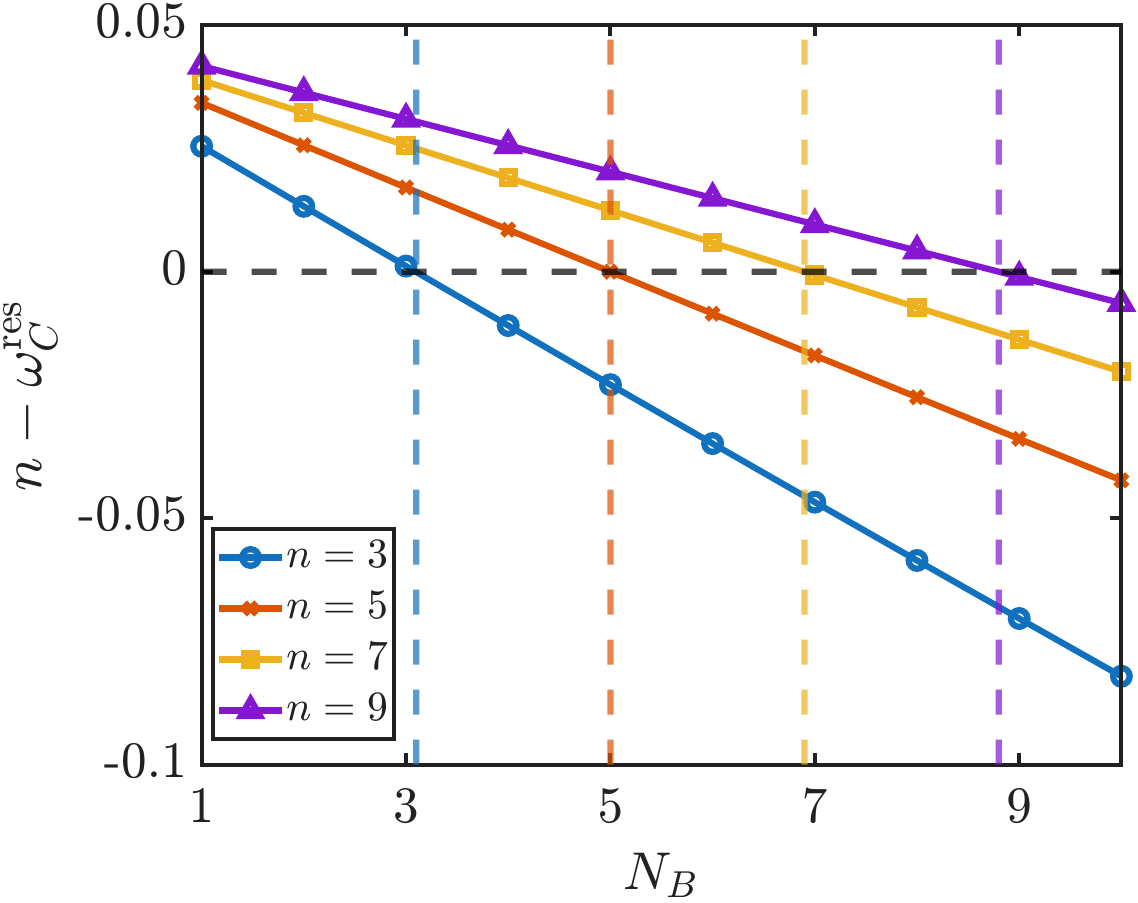}
\caption{Difference between the bare resonance frequency and the true resonance frequency as a function of the particle number in the battery, for $g_{BC}=0.1$. Each colored line with markers corresponds to a different value of $n$. The horizontal black dashed line indicates zero shift, while the colored dashed lines mark the values of $N_B$ at which the shift vanishes for each $n$.}
\label{fig: shift vs NB}
\end{figure}

Figure~\ref{fig: shift vs NB} shows the difference between the bare resonance frequency and the true resonance frequency $\omega_C^{\text{res}}$ as a function of the particle number $N_B$. For particle numbers smaller than the target excitation level ($N_B<n$), the resonance frequency lies below the bare value, $\omega_C^{\mathrm{res}}<n$. This behavior can be understood by comparing the interaction energies of the initial and charged states. For $N_B<n$, the battery--charger interaction energy is larger in the initial state than in the charged state. Consequently, part of the excitation energy required to reach the charged state is already provided by the interaction, and the charger frequency must be reduced below the bare resonance value to restore resonance. As the particle number increases, the magnitude of the shift decreases and becomes nearly zero around $N_B=n$. This explains why the bare resonance already performs optimally near $N_B=n$, as observed in Fig.~\ref{fig: dynamics}(a).
For $N_B>n$, the situation reverses and the resonance shifts above the bare value, $\omega_C^{\mathrm{res}}>n$. In this regime, the interaction energy of the charged state exceeds that of the initial state. Additional energy is therefore required to populate the target excited state, and the charger frequency must be increased above the bare resonance value to compensate for this interaction-induced energy difference. The interaction-induced modification of the spectrum therefore grows with particle number and increasingly detunes the system from the bare resonance. Consequently, the charging efficiency decreases unless the charger frequency is adjusted accordingly.

These results demonstrate that the optimal charger frequency depends not only on the target excitation level but also on the battery size. By tuning $\omega_C$ to the interaction-shifted resonance, one can restore the maximum-energy-transfer condition even when the bare resonance fails. The charging dynamics at the true resonance are shown in Fig.~\ref{fig: dynamics}(e)--(h), where $\omega_C=\omega_C^{\mathrm{res}}$ is individually found for each battery size $N_B$. As predicted, the true resonance allows complete energy transfer from the charger to the battery regardless of the particle number. At the optimal charging time, there is also no gap between the stored work and the ergotropy. This occurs because the entropy vanishes exactly at $t_{\text{opt}}$, indicating the absence of correlations between the battery and the charger, both returning to pure states after the total energy transfer. 

Similar to the bare-resonance case, increasing the particle number reduces the optimal charging time. Again, this can be attributed to the increased interaction energy when the battery has more particles. However, when in true resonance the interaction energy oscillates with a period $2\,t_{\text{opt}}$ (except for the case of $N_B=n=3$), which is the same oscillation frequency as both the stored work $W_B$ and the switching cost $W_{\text{sw}}$. This leads to a significant change in interaction energy between the beginning and the end of the charging process results in a notably larger switching cost than in the bare-resonance case. For $N_B<3$, the instant at which the battery receives the maximum energy coincides with the maximum of $W_{\text{sw}}$, which is positive. The switching cost decreases with increasing particle number, indicating that larger batteries require a smaller charging overhead. This trend continues until $N_B=n=3$, where the switching cost is nearly zero at the optimal charging time. For $N_B>3$, the switching cost becomes negative at $t_{\mathrm{opt}}$, corresponding to a minimum of $W_{\text{sw}}$.

To understand this behavior, we decompose the switching cost into two contributions,
\begin{equation} \label{eq: Wsw decomposition}
    W_{\text{sw}}= W_{\text{shift}} + W_{\text{ex}},
\end{equation}
where $W_{\text{shift}}$ is the energy required to shift the charger frequency from the bare-resonance value to the true resonance, while $W_{\text{ex}}$ represents the excitation energy generated during the interaction quench.
The resonance shift originates from the interaction energy between the battery and the charger, which modifies the energy difference between the initial and charged states. 
From the effective two-level model Eq.~\eqref{Eq:eff_ham}, we can define the shift as the difference between the bare energies of $|0\rangle$ and $|1\rangle$
\begin{align} \label{eq: W shift}
    W_{\text{shift}}
    &= n - \omega_C^{\text{res}} \nonumber\\
    &= \frac{g_{BC}}{(1+\omega_C)^{n+1/2}}
    \sqrt{\frac{\omega_C}{\pi}}
    \left[
        P_n(\omega_C)
        - N_B(1+\omega_C)^{n-1}
    \right],
\end{align}
and which is zero at the bare-resonance as $\omega_C=n$.

As discussed above in Fig.~\ref{fig: shift vs NB}, for $N_B<n$, the true resonance frequency satisfies $\omega_C^{\text{res}}<n$, resulting in a positive shifting cost. Physically, this means that additional energy must be put into the total battery--charger system by the external driving during the quench, i.e. $\text{Tr}\left[\left(\rho^{BC}(t)-\rho^{BC}(0) \right)H_0^{BC}\right]>0$, in order to compensate for the interaction-induced shift and bring the system into resonance. As $N_B$ increases, the magnitude of the shift decreases and eventually almost vanishes when $N_B=n$.
For $N_B>n$, the situation reverses ($\omega_C^{\mathrm{res}}>n$), leading to a negative shifting cost whose magnitude increases with $N_B$. 
Physically, a negative shifting cost indicates that the total energy of the battery--charger system decreases during the charging process, i.e. $\text{Tr}\left[\left(\rho^{BC}(t)-\rho^{BC}(0) \right)H_0^{BC}\right]<0$. Since the battery--charger system is otherwise isolated, the only energy exchange possible is with the external control responsible for switching the interaction. Consequently, maintaining the resonance condition requires energy to be extracted from the total battery--charger system by the external driving field. 
Therefore, by controlling the battery size $N_B$, the coupling strength $g_{BC}$, and the charger frequency $\omega_C$, one can selectively excite a desired battery state while maintaining the resonance condition required for maximum energy transfer. However, this does come at the expense of a finite energy cost which is is required to shift the total system into resonance.

\begin{figure}
\center
\includegraphics[width=\columnwidth]{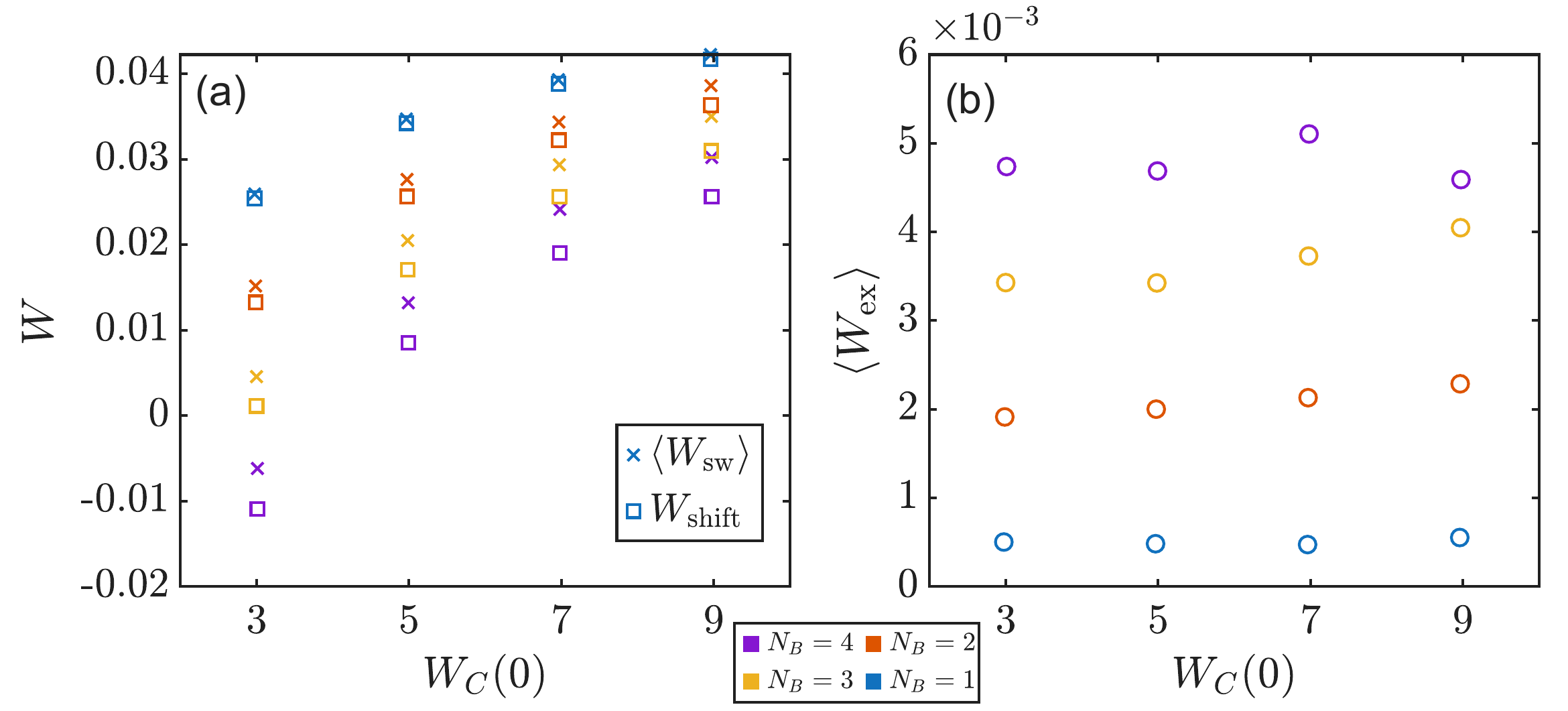}
\caption{(a) Shifting cost $W_{\text{shift}}$ (squares) and average switching cost $\langle W_{\text{sw}}\rangle$ (crosses) as functions of the initial charger energy $W_C(0)$ for different particle numbers $N_B$. (b) Average excitation energy generated during the charging process, obtained from $\langle W_{\text{ex} \rangle}=\langle W_{\text{sw}}\rangle-W_{\text{shift}}$. The coupling is fixed at $g_{BC} =0.1$ for all data. The blue markers correspond to the single-particle battery, while the red, yellow and purple markers represent non-interacting two-, three- and four-particle batteries, respectively.}
\label{fig: shift vs WC}
\end{figure}

In Fig.~\ref{fig: shift vs WC}(a) we compare the shifting cost and the total switching cost as functions of the charger's initial energy, $W_C(0)=\omega_C^{\text{res}}$, for up to $N_B=4$ particles. Due to the fast oscillations of the breathing mode, we compute the average of the switching cost over $\left[t_\mathrm{opt}-\dfrac{\pi}{2},t_\mathrm{opt}+\dfrac{\pi}{2} \right]$. As expected, the shifting cost decreases as $N_B$ approaches $n$ and eventually becomes negative for $N_B>n$. For a fixed particle number, larger charger energies generally require a larger switching cost because the resonance shift becomes more pronounced.
To quantify the amount of excitations generated during the charging process at the true resonance, we subtract the shifting cost from the total switching cost and define the average contribution purely from non-equilibrium excitations $\langle W_{\text{ex}} \rangle$. These manifest as the charger and battery states being excited to other eigenstates of the decoupled Hamiltonians $H^C$ and $H^B$ respectively, therefore representing a leakage of energy to states which cannot be  extracted later. The resulting values are shown in Fig.~\ref{fig: shift vs WC}(b). As expected, the excitation energy increases with particle number due to the larger number of available excitation channels in larger many-particle systems. In contrast, increasing the initial energy of the charger does not necessarily lead to larger excitation production. Notably, $\langle W_{\text{ex}} \rangle$ is typically more than one order of magnitude smaller than $W_{\text{shift}}$ and remains negligible compared to the available charger energy $W_C(0)$ throughout the weak-coupling regime, when $g_{BC}\ll1$. However, this can become significant when the coupling increases, as we will discuss in Section \ref{subsection: stronger quench}.

\subsection{Power enhancement and Quantum speed limit time}
Next, let us focus on the charging power of the few-body batteries and how this can be optimized. The charging power depends on how much energy can be transferred, along with the speed of this process. The minimum time to connect two quantum states can be well defined by so called quantum speed limit (QSL) times. Using the expression for the optimal charging time Eq.~\eqref{eq: optimum time} for the time-evolved state Eq.~\eqref{eq: psi_t}, one finds
\begin{equation}
    |\Psi(t_\mathrm{opt})\rangle = \dfrac{\delta}{\Omega}|0\rangle+\dfrac{2J}{\Omega}|1\rangle.
\end{equation}
For our purposes, we are interested in the evolution between the initial state $|\Psi(0)\rangle=|0\rangle$ and the optimally charged state $|\Psi(t_\mathrm{opt})\rangle$. The corresponding generalized Mandelstam--Tamm QSL time is given by~\cite{Mandelstam1945,PhysRevLett.124.110601}
\begin{equation} \label{eq: definition speed limit}
    \tau_{\mathrm{QSL}} = \frac{\arccos{|\langle \Psi(0)|\Psi(t_\mathrm{opt}) \rangle|}}{\sqrt{\langle\Psi(0)|\mathcal{H}^2|\Psi(0)\rangle - \langle\Psi(0)|\mathcal{H}|\Psi(0)\rangle^2}}\,,
\end{equation}
where $\arccos{|\langle \Psi(0)|\Psi(t_\mathrm{opt}) \rangle|}$ is the Bures angle between the initial and optimally charged states.
Using the effective two-level Hamiltonian $\mathcal{H}=H_1^{BC}$ derived above, we obtain
\begin{equation} 
    \tau_{\mathrm{QSL}}=\dfrac{1}{|J|} \arccos\left({\dfrac{|\delta|}{\sqrt{(2J)^2+\delta^2}}}\right).
\end{equation}
One can readily verify that the optimal charging time is bounded from below by the QSL,
\begin{equation}
    t_\mathrm{opt} \geq \tau_{\mathrm{QSL}},
\end{equation}
with equality achieved only at the true resonance condition, $\delta=0$. In this case, the initial and final states become orthogonal and the generalized Mandelstam--Tamm bound reduces to
\begin{equation} \label{eq: tqsl}
    \tau_{\text{QSL}} = \frac{\pi}{2 g_{BC} \sqrt{N_B} |\mathcal{I}_n|}.
\end{equation}
Therefore, as long as the battery is charged in the weak-coupling regime, where the two-level approximation remains valid, and the charger frequency is tuned exactly to resonance, Eq.~\eqref{eq: tqsl} provides a tight bound for the optimal charging time required to achieve maximum energy transfer. 

The corresponding charging power at the optimal charging time is then given by
\begin{equation} \label{eq: TLM-power}
    \mathcal{P}_W(t_{\text{opt}})
    = \frac{W_C(0)}{\tau_{\text{QSL}}}
    = \frac{2 W_C(0) g_{BC}\sqrt{N_B}|\mathcal{I}_n|}{\pi}.
\end{equation}
Equation~\eqref{eq: TLM-power} suggests two possible routes to enhance the charging power. The first is to increase the inter-species coupling strength $g_{BC}$ between the battery and the charger. However, a strong interaction quench can lead to more complicated dynamics due to the generation of higher-energy excitations and irreversible work, as discussed in the previous section. In this regime, the two-level approximation breaks down because particles in the battery can populate excited states outside the effective two-level subspace. Moreover, the production of irreversible work can contribute to the energy of both the battery and the charger, making it difficult to unambiguously quantify the amount of energy transferred between the two subsystems.
The second route is to increase the number of particles in the battery while keeping the coupling strength fixed. In this case, the charging power scales as $\sqrt{N_B}$, providing a genuine many-body enhancement without requiring stronger interactions, which are also observed in other battery models \cite{PhysRevLett.120.117702, PhysRevLett.118.150601, PhysRevB.98.205423}. This scaling will be verified numerically in the following section.

\begin{figure}
\center
\includegraphics[width=\columnwidth]{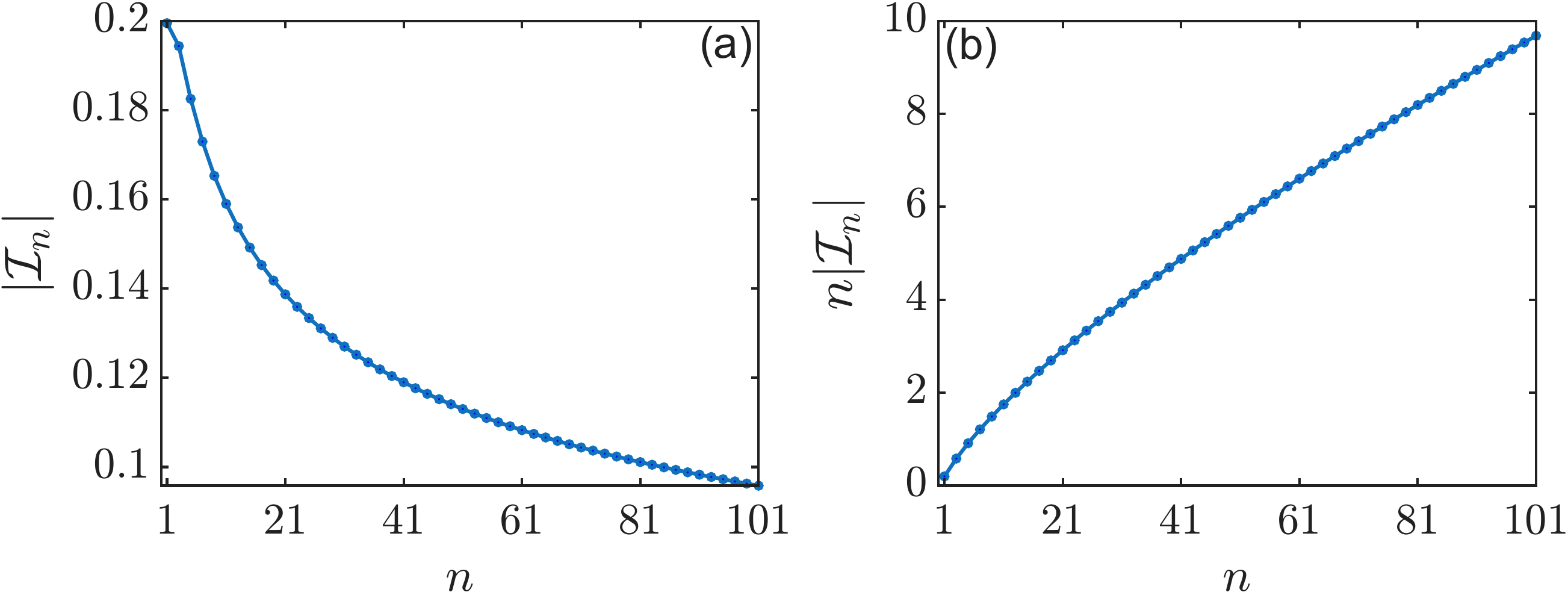}
\caption{(a) The value of $|\mathcal{I}_n|$ and (b) $n|\mathcal{I}_n|$ as functions of $n$.}
\label{fig: In}
\end{figure}

Lastly, we must consider the overlap coefficient $\mathcal{I}_n$, which connects the initial state of the empty battery and charged charger, to the final state in which the battery is charged to the $n^{\text{th}}$ level and charger is discharged. For odd values $n = 2k + 1$, the overlap coefficient $\mathcal{I}_n$ takes the analytical form
\begin{equation}
    \mathcal{I}_n = (-1)^{(n-1)/2}
    \frac{1}{\left(\frac{n-1}{2}\right)!}
    \sqrt{\frac{n!}{2^{\,n-1}\pi}}
    \frac{\omega_B\,\omega_C^{(n+1)/2}}
         {(\omega_B+\omega_C)^{(n+2)/2}}.
\end{equation}
In the weak-coupling regime, the resonance condition implies $W_C(0)\simeq n$. Setting $\omega_B = 1$, we find that $|\mathcal{I}_n|$ decreases with increasing $n$, as shown in Fig.~\ref{fig: In}(a). Consequently, the time required to reach maximum energy transfer becomes longer for higher charger energies. Nevertheless, the decrease of $|\mathcal{I}_n|$ with $n$ is relatively slow. Since the charging power scales as $n|\mathcal{I}_n|$, the overall charging power still increases with $n$, although sublinearly, as illustrated in Fig.~\ref{fig: In}(b). This behavior can be understood physically from the decreasing overlap between the charger state and higher excited states of the battery. Excitations to higher-energy levels are less effectively driven by a weak interaction, resulting in a smaller interaction matrix element and therefore a reduced charging rate. One way to compensate for this reduction is to increase the coupling strength $g_{BC}$, thereby maintaining a sufficiently large interaction energy across different excitation levels, provided that the accompanying increase in the switching cost remains low. This will be discussed in more detail in the next section.
 
\begin{figure}
\center
\includegraphics[width=\columnwidth]{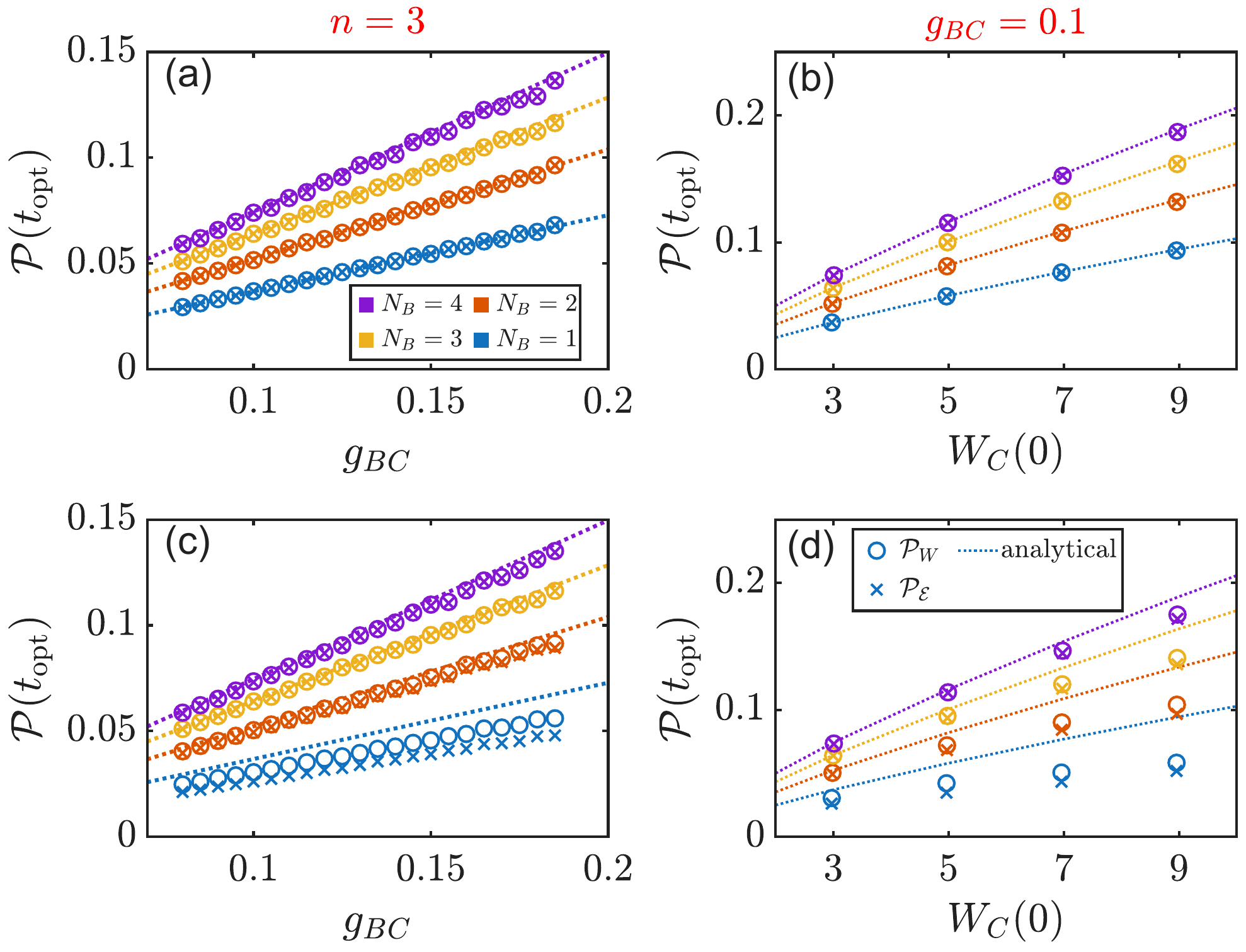} 
\caption{(a) Charging power and ergotropic power evaluated at the optimal charging time, plotted as functions of the inter-species coupling strength $g_{BC}$ at the resonance frequency $\omega_C^\text{res}$ for $n=3$. (b) Charging power and ergotropic power as functions of the initial charger energy $W_C(0)$ for $g_{BC}=0.1$. (c) and (d) Same as (a) and (b), respectively, but for the bare-resonance condition $\omega_C=3$. Circles and crosses denote numerical results obtained from exact diagonalization for the charging power and ergotropic power, respectively, while the dotted lines represent the analytical prediction of the charging power from the two-level model [Eq.~\eqref{eq: TLM-power}]. In (b) and (d), the dotted lines connecting the analytical results for odd values of $W_C(0)$ serve as guides to the eye. In all panels, the blue curves correspond to the single-particle battery, while the red, yellow and purple curves represent non-interacting two-, three- and four-particle batteries, respectively.}
\label{fig: power}
\end{figure}

\begin{figure}
\center
\includegraphics[width=\columnwidth]{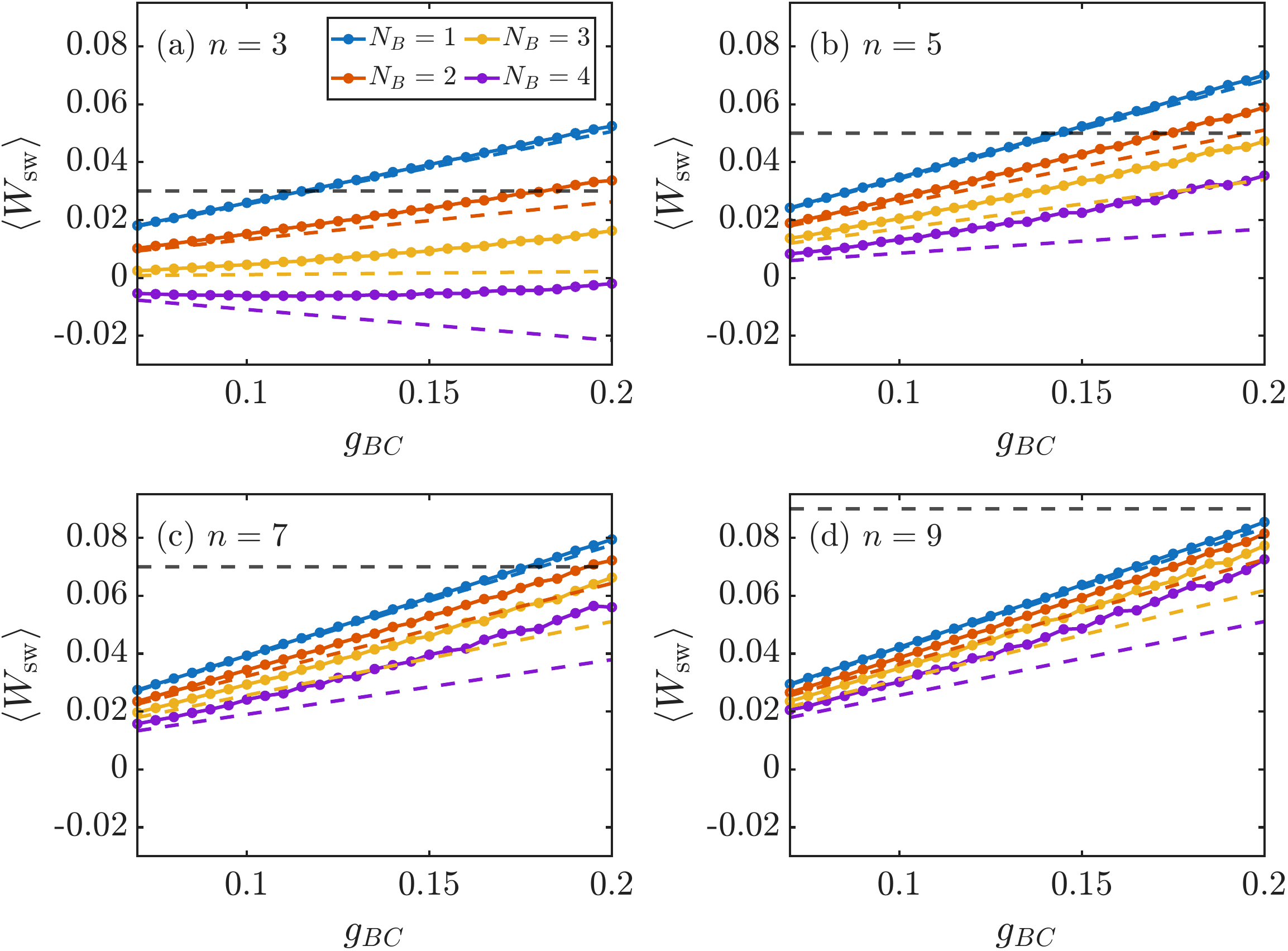}
\caption{(a)–(d) The average switching cost generated during the charging process, for different values of $n$ at the true resonance. The black dashed line indicates one percent of the initial charger energy. The blue solid lines correspond to the single-particle battery, while the red, yellow and purple solid curves represent non-interacting two-, three- and four-particle batteries, respectively. The corresponding colored dashed lines denote the shifting cost $W_{\text{shift}}$ defined in Eq.~\eqref{eq: W shift}.}
\label{fig: W switching}
\end{figure}

To validate these predictions, we compare the performance of the non-interacting many-body bosonic quantum battery with that of its single-particle counterpart. Figures~\ref{fig: power}(a) and (b) show the charging performance when the battery and charger are tuned to the true resonance, while Figs.~\ref{fig: power}(c) and (d) correspond to the bare-resonance condition.
As shown in Fig.~\ref{fig: power}(a), tuning the charger to the true resonance $\omega_C^\text{res}$ results in a charging power that increases linearly with the interaction strength $g_{BC}$. Moreover, increasing the number of particles in the battery leads to a collective enhancement of the charging power compared with the single-particle case. For larger values of $g_{BC}$, the numerical results obtained from exact diagonalization (circles) begin to deviate slightly from the analytical prediction of the two-level model (dotted lines). This discrepancy originates from the generation of irreversible excitations at stronger couplings and larger particle numbers, as will be discussed later in Fig.~\ref{fig: W switching}.

In addition to the charging power, we also consider the ergotropic power,
\begin{equation}
    \mathcal{P}_{\mathcal{E}}(t_{\text{opt}})
    = \frac{\mathcal{E}(t_{\text{opt}})}{t_{\text{opt}}},
\end{equation}
which quantifies the rate at which extractable work is transferred from the charger to the battery. As expected from the two-level model, the ergotropic power is nearly identical to the charging power at the true resonance, indicating that essentially all the transferred energy remains extractable.
Figure~\ref{fig: power}(b) shows the charging and ergotropic power as functions of the initial energy stored in the charger. Larger initial charger energies generally lead to higher charging power. Although the power increases approximately linearly with $W_C(0)$, a slight sublinear behavior is observed due to the gradual decrease of the overlap $|\mathcal{I}_n|$, as discussed in the previous section.

A markedly different behavior is observed when the charger frequency is fixed at the bare-resonance value rather than the true resonance. As shown in Figs.~\ref{fig: power}(c) and (d), both the charging power and the ergotropic power are reduced compared with the resonant case. The degradation is particularly pronounced for the ergotropic power, which exhibits a visible gap from the charging power. Furthermore, the difference between the true-resonance and bare-resonance results becomes more significant as the interaction strength increases. This can be understood from the interaction-induced shift of the resonance frequency: stronger coupling pushes the true resonance further away from the bare value, thereby reducing the efficiency of the charging protocol when the charger frequency is kept fixed.

A similar effect occurs when the particle number is held fixed and the charger is prepared with larger initial energy, see Figs.~\ref{fig: power}(d). Since the resonance shift grows with the excitation energy, operating at the bare resonance increasingly deviates from the optimal charging condition. Consequently, both the charging and ergotropic power become progressively reduced.
Overall, the charging power at the true resonance provides an upper bound on the achievable charging and ergotropic power within the weak-coupling regime, corresponding to the idealized situation in which the battery is charged exactly at the QSL time. 

To further understand the scaling of the switching cost in many-body batteries
, we examine this quantity as a function of $g_{BC}$ in Fig.~\ref{fig: W switching} at the true resonance frequency.
To excite a particle in the battery to the third excited state ($n=3$) [see Fig.~\ref{fig: W switching}(a)], the required switching cost is larger for a single-particle battery than for its many-body counterparts. This behavior originates primarily from the shifting cost, which becomes significantly larger as the particle number deviates further from the target excitation level $n$. On the other hand, the gap between the solid and dashed curves, corresponding to the average excitation contribution $\langle W_{\text{ex}} \rangle$, increases with particle number, in agreement with the results shown in Fig.~\ref{fig: shift vs WC}. These findings highlight a two-fold advantage of many-body QBs: enhanced charging power and a reduced switching cost, provided that the particle number is chosen close to the target resonance condition ($N_B \simeq n$).
As expected, increasing $g_{BC}$ results in a larger switching cost, exhibiting approximately linear scaling in the weak-coupling regime. The fluctuations observed at stronger couplings for larger particle numbers originate from the more pronounced oscillations in the time evolution of $W_{\text{sw}}$ [see Fig.~\ref{fig: dynamics}(h)].

For larger stored energies, shown in Fig.~\ref{fig: W switching}(b)–(d), the excitation contribution increases only slowly with $n$. Moreover, the switching cost remains significantly smaller than the initial charger energy throughout the parameter range considered. The black dashed line indicates 1\% of the charger energy, and we find that for $N_B=4$, the switching cost remains below this threshold for couplings up to approximately $g_{BC}=0.2$, even for $n=9$. Furthermore, the switching costs of larger batteries become comparable to those of the single-particle battery as the stored energy increases. This relatively weak growth of excitation production allows one to employ stronger inter-species interactions $g_{BC}$ to enhance energy transfer without generating excessive excitations in the system.
It is also worth noting that, in the stronger-coupling regime, the gap between the total switching cost and the shifting cost becomes increasingly pronounced, indicating that a larger fraction of the supplied energy is converted into excitations during the charging process. This behavior reflects the gradual breakdown of the two-level approximation as higher excited states become populated.

Finally, let us comment on the role of particle statistics in the charging performance. We can compute the QSL time when the battery consists of $N_B$ non-interacting fermions as
\begin{equation}
   \tau^F_{\text{QSL}} = \frac{\pi}{2 g_{BC} |\mathcal{I}^F_n|},
\end{equation}
where the overlap factor is given by
\begin{equation}
    \mathcal{I}^F_n = \int \phi^*_{N_B-1}(x)\,\varphi_1^*(x)\,\phi_{N_B+n-1}(x)\,\varphi_0(x)\,dx.
\end{equation}
This behavior can be understood in terms of Fermi blocking. Due to the Pauli exclusion principle, fermions are forbidden from occupying the same single-particle orbital. As a result, only the particle at the surface of the Fermi sea, occupying the $(N_B-1)$th orbital, can participate in the charging process and be excited to the $(N_B+n-1)$th orbital. Consequently, the collective bosonic enhancement present in the bosonic battery, where multiple particles coherently contribute to the charging dynamics through symmetric occupation of low-energy states, is entirely absent. The charging process is therefore limited to a single effective fermionic transition, and the charging power no longer exhibits any scaling with the particle number $N_B$.

Moreover, as $N_B$ increases, the overlap $\mathcal{I}^F_n$ further decreases. This suppression arises because the charger wave functions $\varphi_0(x)$ and $\varphi_1(x)$ are strongly localized near the center of the harmonic trap, while the higher-energy fermionic orbitals $\phi_{N_B-1}(x)$ and $\phi_{N_B+n-1}(x)$ become increasingly delocalized and oscillate rapidly over a much larger spatial region. The resulting spatial mismatch strongly suppresses the overlap between the initial and charged states, leading to a rapid decay of $|\mathcal{I}^F_n|$ with increasing $N_B$.
As a consequence, the quantum speed-limit time $\tau^F_{\text{QSL}}$ actually increases with particle number, further degrading the charging performance. The fermionic battery thus behaves effectively as a collection of independent single-particle modes, for which collective speedup and power enhancement are intrinsically forbidden, resulting in substantially reduced charging efficiency compared to the bosonic case.

\subsection{Optimize the charging power under a total-cost constraint}
As discussed in the previous section, the reduction of the switching cost with increasing particle number $N_B$ allows one to employ stronger inter-species interactions $g_{BC}$ to enhance the charging performance without generating excessive excitations. This observation motivates the following optimization strategy: instead of fixing the coupling strength, we choose $g_{BC}$ such that the total charging cost, defined below, remains below a prescribed threshold,
\begin{equation}
    \langle C_\mathrm{total}\rangle = \langle W_\mathrm{ex} \rangle  + |W_\mathrm{shift}|,
\end{equation}
where $W_{\mathrm{ex}}$ and $W_{\mathrm{shift}}$ are given by Eqs.~\eqref{eq: Wsw decomposition} and~\eqref{eq: W shift}, respectively. The modulus is introduced because the shifting cost can become negative when $\omega_C^{\mathrm{res}}>n$, as shown in Figs.~\ref{fig: shift vs WC}(a) and~\ref{fig: W switching}(a). In this regime, the charger must initially be prepared with excess energy, and part of this energy is subsequently released to the external driving field during the quench in order to bring the system into resonance. We treat this as an energy cost as this excess energy is not extracted as usable work, but rather it is lost to the magnetic field which controls the short range interaction strength. 

\begin{figure*}
\center
\includegraphics[width=2\columnwidth]{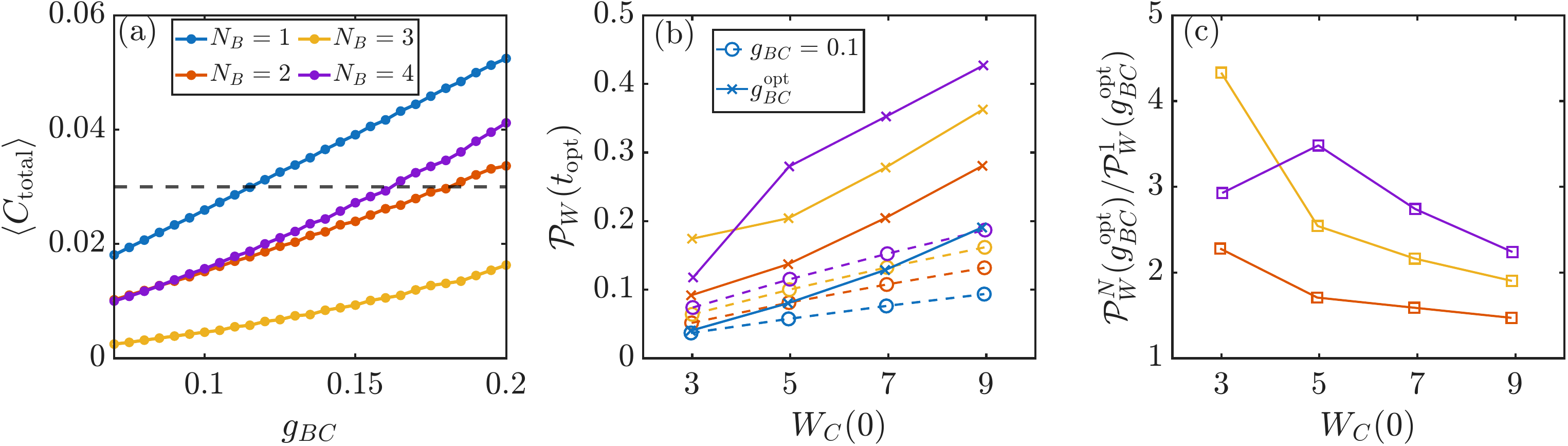}
\caption{(a) Total charging cost $\langle C_{\mathrm{total}} \rangle$ as a function of the battery--charger coupling strength $g_{BC}$ for different particle numbers and $n=3$. (b) Charging power as a function of the initial charger energy. The dashed lines with circles show the charging power at the true resonance for a fixed coupling strength $g_{BC}=0.1$ (same data as in Fig.~\ref{fig: power}(b)), while the solid lines with crosses correspond to the optimized charging power obtained by choosing $g_{BC}^\mathrm{opt}$ such that the total average cost remains below $1\%$ of the initial energy stored in the charger. (c) Ratio between the optimized charging power of the many-body batteries and that of the corresponding single-particle battery. In all panels, the blue curves correspond to the single-particle $N_B=1$ battery, while the red, yellow and purple curves represent non-interacting two-, three- and four-particle batteries, respectively.}
\label{fig: optimize}
\end{figure*}

Fig.~\ref{fig: optimize}(a) shows the total charging cost for $n=3$ for up to $N_B=4$ particles. The total cost is the same as $\langle W_\mathrm{sw} \rangle$ for $N_B =1, 2, 3$. However, for $N_B =4$, the average charging cost now is positive and comparable to the $N_B=2$ case. This can be explained as in the weak coupling regime, most of the cost is the shifting cost, which should be almost equivalent for $N_B=2$ and $N_B=4$ to bring it into resonance with respect to $n=3$. For larger interactions, more excess excitations $\langle W_{\text{ex}}\rangle$ are created in bigger batteries, which makes the total cost for $N_B=4$ higher than $N_B=2$. It is worth noticing that for a higher target excitation level, the total cost is the same as the switching cost as shown in Fig.~\ref{fig: W switching}(b)-(d)

We determine the optimal coupling strength $g_{BC}^\mathrm{opt}$ for each value of $n$ by imposing the constraint that the average charging cost remains below $1\%$ of the initial charger energy, as suggested by the results of Fig.~\ref{fig: W switching}(b)--(d) and Fig.~\ref{fig: optimize}(a). In Fig.~\ref{fig: optimize}(b) we show the charging power as a function of the initial charge $W_C(0)$ using the optimized coupling $g_{BC}^{\text{opt}}$, and we also compare this to keeping it fixed at $g_{BC}=0.1$. Since the switching cost decreases as $n$ increases, substantially stronger interactions can be employed while still satisfying the cost constraint. As a result, the charging power can be significantly enhanced compared to the fixed-coupling scenario. In particular, the charging power no longer displays a sublinear growth with respect to $W_C(0)$, showing that powerful larger capacity batteries can also be realized when the coupling is optimized.  

The enhancement is particularly pronounced when the particle number is close to the target excitation level. To quantify this effect, Fig.~\ref{fig: optimize}(c) shows the ratio between the optimized charging power of the many-body batteries and that of the corresponding single-particle battery, $\mathcal{P}^N_W(g_{BC}^{\text{opt}})/\mathcal{P}^1_W(g_{BC}^{\text{opt}})$. For $n=3$, operating the battery with $N_B=3$ minimizes the resonance shift and therefore allows access to considerably stronger coupling strengths while remaining below the prescribed cost threshold. As a result, the charging power can exceed that of the single-particle battery by more than a factor of $N_B$, exhibiting superextensive scaling when at the many-body resonance. In contrast, when the particle number differs substantially from $n$, the larger resonance shift increases the total cost and reduces the achievable enhancement. 
As the initial charger energy increases, this enhancement gradually decreases and eventually approaches the characteristic $\sqrt{N_B}$ scaling. This behavior can be understood from Fig.~\ref{fig: W switching}(d). For larger values of $n$, the difference in switching cost between batteries with different particle numbers becomes smaller. Consequently, the optimized coupling strengths satisfying the switching-cost constraint become nearly identical for different values of $N_B$, particularly when $N_B \ll n$. In this limit, the optimization no longer provides an additional advantage beyond the many-body enhancement predicted by the two-level model, with the charging power approaching the QSL scaling given by Eq.~\eqref{eq: TLM-power}.

\subsection{Stronger interaction quench} \label{subsection: stronger quench}
\begin{figure}
\center
\includegraphics[width=\columnwidth]{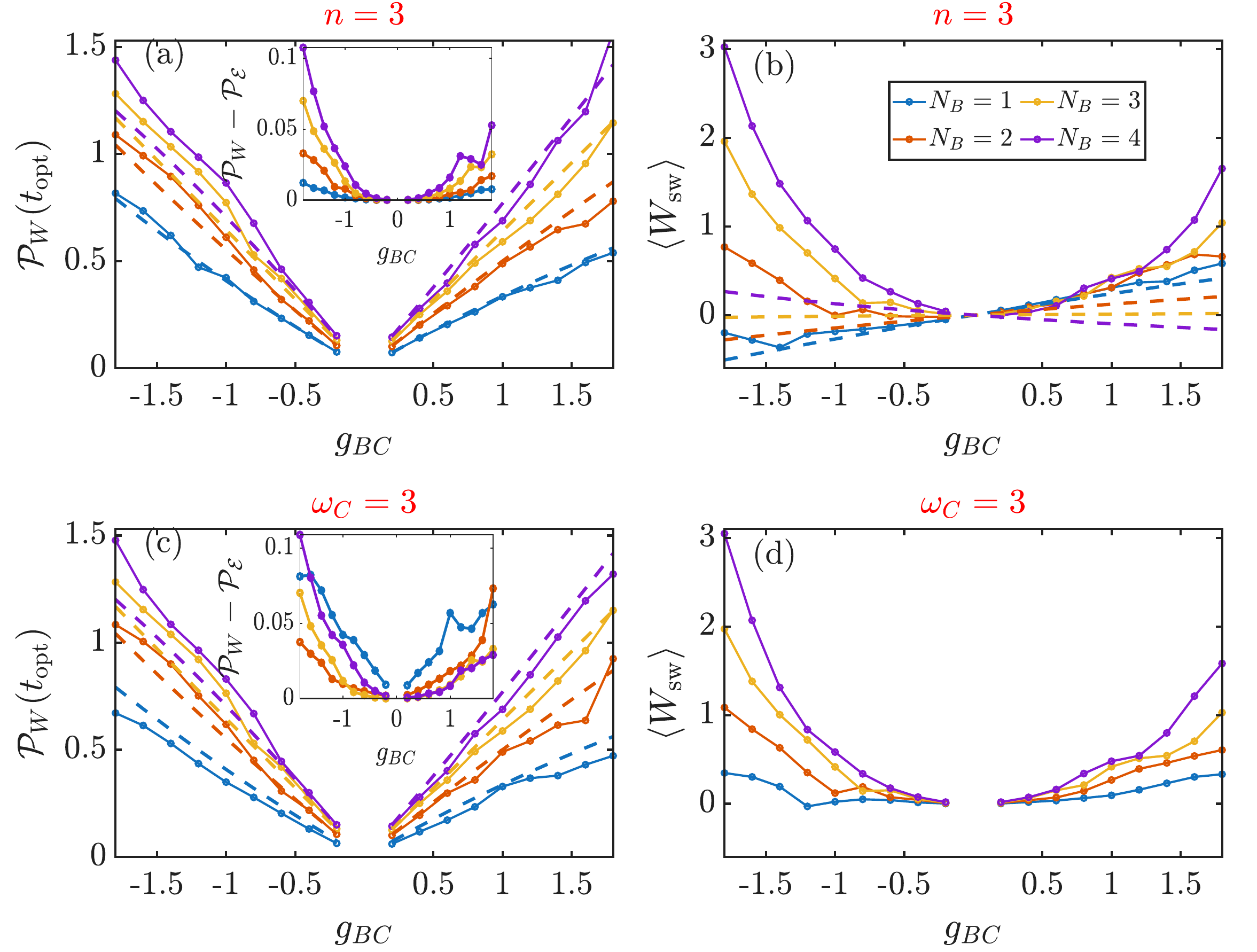} 
\caption{(a)--(b) Charging power and average switching cost evaluated at the optimal charging time, plotted as functions of the inter-species coupling strength $g_{BC}$ at the true resonance frequency $\omega_C^\mathrm{res}$ for $n=3$ in the stronger-coupling regime. (c)--(d) Same as (a)--(b), but for the bare resonance condition $\omega_C=3$. In panels (a) and (c), the dashed lines represent the analytical prediction of the charging power obtained from the two-level model. The insets display the difference between the charging power and the ergotropic power as a function of $g_{BC}$. In panel (b), the dashed lines correspond to the shifting cost associated with the resonance-frequency shift. In all panels, the blue curves correspond to the single-particle battery, while the red, yellow and purple curves represent non-interacting two-, three- and four-particle batteries, respectively.}
\label{fig: P Wirr stronger gBC}
\end{figure}

For stronger battery--charger couplings, the linear scaling of both the charging power and the switching cost with $g_{BC}$ breaks down, as shown in Figs.~\ref{fig: P Wirr stronger gBC}(a) and (b), even when the charger frequency is tuned to the true resonance. As the magnitude of the coupling increases, the shifting cost grows more rapidly, both for attractive and repulsive interactions $g_{BC}$, since stronger coupling shifts the resonance frequency further away from the bare resonance value. In addition, the gap between the total switching cost and the shifting cost becomes substantially larger in the strong-coupling regime, indicating an increased population of higher excited states during the charging process. Larger particle numbers further increase $\langle W_{\text{sw}} \rangle$, reflecting the larger number of excitation channels available in the many-body system.

As a result, the charging dynamics can no longer be described by a single resonant transition between the initial and charged states. The impact of these excitations on the charging power, however, differs between the repulsive $g_{BC}>0$ and attractive $g_{BC}<0$ sides. In the strongly repulsive regime, the charging power evaluated at the optimal charging time exhibits a clear sublinear dependence on $g_{BC}$. Although stronger interactions increase the characteristic energy scales of the system, the population of higher excited states causes the optimal charging time to become longer than the QSL time predicted by the two-level model. Consequently, the charging advantage predicted by the weak-coupling theory is gradually lost and the charging power falls below the two-level prediction.

In contrast, for sufficiently strong attractive couplings and larger particle numbers, both the charging and ergotropic power can exceed the linear trend predicted by the weak-coupling theory. While the two-level approximation also breaks down in this regime, the optimal charging time becomes shorter than the QSL time obtained from the effective two-level model. Together with the additional energy injected into the system during the quench, this leads to charging powers that can exceed the two-level prediction despite the larger excitation cost. 

Another signature of the strong-coupling regime is the growing difference between the charging power and the ergotropic power, as shown in the inset of Fig.~\ref{fig: P Wirr stronger gBC}(a). As the coupling strength increases, the larger population of higher excited states generates a more mixed battery state and stronger battery--charger correlations. Consequently, an increasing fraction of the stored energy becomes inaccessible for work extraction, leading to a larger gap between the charging power and the ergotropic power.

Similarly, Figs.~\ref{fig: P Wirr stronger gBC}(c) and (d) show the corresponding results obtained at the bare resonance. For the single-particle battery, the gap between the charging power and the ergotropic power is significantly larger than in the true-resonance case because the interaction-induced shift of the resonance frequency moves the system further away from the optimal operating point. In contrast, for larger particle numbers, the difference between the bare-resonance and true-resonance protocols becomes less pronounced. In this regime, the charging dynamics are dominated by the large amount of excitation generated during the quench, and the imperfections arising from operating away from the true resonance become comparatively less important.

\begin{figure}
\center
\includegraphics[width=0.9\columnwidth]{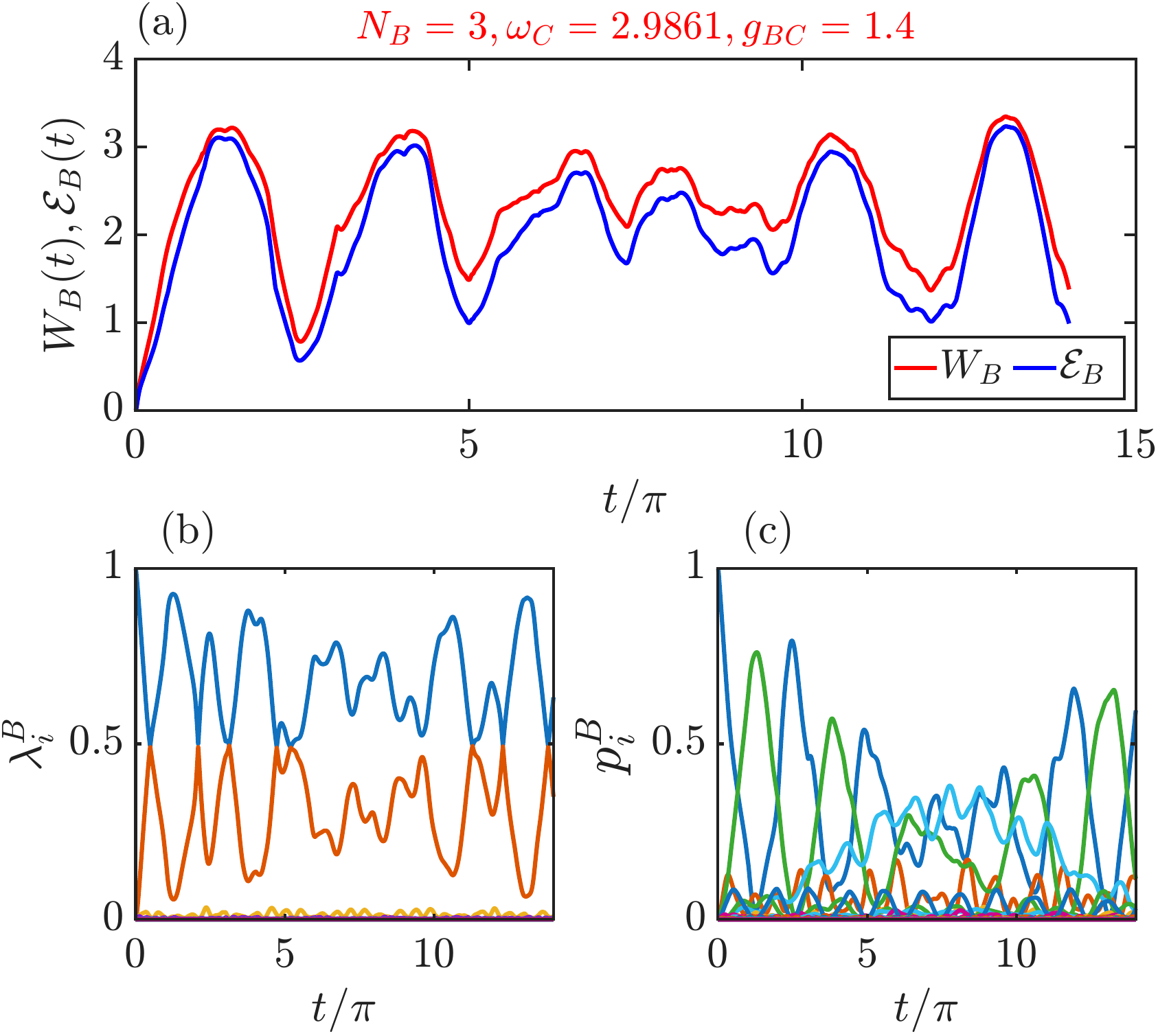} 
\caption{(a) The time-dependent stored work (red) and ergotropy (blue) $N_B=3$ quantum battery. The value of $\omega_C$ is chosen by solving the Eq.~\eqref{eq: delta} for $g_{BC} = 1.4$ and $n=3$. (b) The passive-state probabilities $\lambda_i$ and (c) the battery-state probabilities $p_i$ corresponding to panel (a).} 
\label{fig: stronger g_BC}
\end{figure}

Figure~\ref{fig: stronger g_BC}(a) shows the time-dependent stored work and ergotropy of the $N_B=3$ battery for $g_{BC}=1.4$. The charger frequency is chosen by solving Eq.~\eqref{eq: delta} for $n=3$. Compared to the weak-coupling regime [cf. Fig.~\ref{fig: dynamics}(e)], the time required to reach the first maximum of stored work is significantly reduced. At the same time, the charging dynamics become considerably more complex, as neither the stored work nor the ergotropy exhibits the simple Rabi-like oscillations predicted by the two-level model.

This increased complexity is further illustrated in Figs.~\ref{fig: stronger g_BC}(b) and (c), where the passive-state and battery-state probabilities reveal that several excited states become significantly populated, especially at longer charging times. As the charging duration increases, more excitations are generated during the quench, leading to a progressively mixed battery state and a growing difference between the stored work and the ergotropy. Nevertheless, despite the breakdown of the two-level description, tuning the charger frequency according to the resonance condition of Eq.~\eqref{eq: delta} still enables substantial energy transfer and near-optimal ergotropy in the strong-coupling regime.

\section{Conclusions} \label{sec: conclusions}
In this work, we have investigated the charging dynamics and ergotropy of bosonic many-body QBs and single particles chargers confined in one-dimensional harmonic traps. In the weak-coupling regime, we showed that the battery--charger system can be accurately described by an effective two-level model. Within this framework, analytical expressions for the resonance condition, optimal charging time, QSL, stored work, ergotropy, and charging power were derived. Comparison with exact diagonalization demonstrated excellent agreement, confirming that the two-level description reliably captures the charging dynamics as long as higher-energy excitations remain negligible.

A key result of our work is understanding how the different interaction effects affect the charging efficiency of QBs comprised of ultracold atoms, in particular how collective interactions can boost performance. However, interaction induced energy shifts can also drive the charger and battery out of resonance leading to less energy transfer, while mitigating this effect has the consequence of introducing additional energy costs which need to be accounted for. We have shown how these can be effectively minimized through optimizing the coupling strength, along with exploiting characteristic resonances that emerge due to the interplay between the battery size and charger energy in harmonic oscillator systems. When successfully optimized, the battery can be perfectly charged at the QSL time and show super-extensive scaling with the battery size. While our analysis is predicated on the ability to describe the weakly coupled battery-charger system with an effective two-level model, we show that even in the strong coupling regime the resonant system displays remarkably robust energy transfer and ergotropy. 

Overall, our results reveal how the interplay between resonance engineering, particle number, and interaction strength can be exploited to enhance the performance of many-body QBs comprised of 1D ultracold bosons. The ability to achieve faster charging while maintaining a small switching cost highlights the potential of bosonic many-body systems as scalable quantum energy-storage devices. Given the high degree of control available in ultracold-atom experiments, our model provides a promising route toward the realization and optimization of QBs in realistic settings.

An interesting direction for future work is to investigate the role of finite intra-species interactions within the battery. Since the charging power is ultimately governed by the overlap factor $\mathcal{I}_n$, tuning the interaction strength among battery particles may provide an additional mechanism for controlling the resonance condition and modifying the charging dynamics. In particular, interaction-induced changes to the many-body wave function could be used to enhance the effective overlap between the initial and charged states, potentially leading to shorter QSL times and improved charging performance. Exploring this possibility may open new avenues for interaction-assisted QBs design. 

Furthermore, optimized time-dependent battery--charger interactions could be engineered to simultaneously reduce the charging time and minimize the total switching cost. One promising route is the use of shortcuts to adiabaticity \cite{Santos2019}, which could enable access to the strong-coupling regime while suppressing unwanted excitation generation. The results obtained in the present work suggest that attractive battery--charger interactions may be particularly beneficial in this context. In the strong-coupling regime, we observed charging dynamics that reach the optimal energy-transfer point on timescales shorter than the QSL predicted by the effective two-level model. This indicates that interaction engineering may provide a route to surpassing the charging performance achievable within weak-coupling protocols, allowing faster charging while maintaining efficient energy transfer. Combining such interaction-assisted charging strategies with optimal-control techniques therefore represents a promising direction for realizing high-power QBs.

\section{Acknowledgments}
T.F. would like to thank Nathan Harshman and Steve Campbell for energetic discussions. This work is supported by the Okinawa Institute of Science and Technology Graduate School (OIST). We acknowledge the use of the high-performance computing cluster Deigo provided by the Scientific Computing and Data Analysis section at OIST. This work was partially supported by Japan’s
Council for Science, Technology and Innovation (CSTI) under the Cross-ministerial Strategic Innovation
Promotion Program (SIP) for “Promoting the application for advanced quantum technology platforms
to social issues” (Funding agency: QST, Grant Number JPJ012367) and the JST Grant No. JPMJPF2221. T.F. acknowledges support from JSPS KAKENHI Grant No. JP23K03290.

\appendix 
\section{\label{subsec: ED}Exact Diagonalization}
To study the charging time dynamics of many-body ultracold systems, it is highly advantageous to solve the many-body Hamiltonian in its second quantization form using the Exact Diagonalization method \cite{Zhang_2010}. In this framework, the many-body Hamiltonian can be expressed as 
\begin{align} \label{eq:H0-2ndquantized}
    H^{BC}_0 = \sum_{\sigma \in \{B,C\}}&\sum_{ij} h^{\sigma}_{ij} \hat{a}^\dagger_{\sigma,i}\hat{a}_{\sigma,j} \nonumber\\ 
    &+ \dfrac{1}{2} \sum_{ijk\ell}U^B_{ijk\ell}\hat{a}^\dagger_{B,i}\hat{a}^\dagger_{B,j}\hat{a}_{B,\ell}\hat{a}_{B,k}, 
\end{align}
\begin{equation} \label{eq:H1-2ndquantized}
    H_1^{BC} = H_0^{BC} + \sum_{ijk\ell}U^{BC}_{ijk\ell}\hat{a}^\dagger_{B,i}\hat{a}^\dagger_{C,j}\hat{a}_{C,\ell}\hat{a}_{B,k},
\end{equation}
where 
\begin{align}
    h^\sigma_{ij} &= \int \psi_{\sigma,i}^*(x_\sigma)\hat{H}^\sigma_{\text{sp}} \psi_{\sigma,j}(x_\sigma)dx_\sigma,
\end{align}
are the one-body integrals, with a complete set of single-particle states $|\psi_{\sigma,i}(x_\sigma)\rangle$ can be chosen as the eigenfunctions of the Hamiltonian of a harmonic oscillator $\hat{H}^\sigma_{\text{sp}} = -\dfrac{\hbar^2}{2m_\sigma} \dfrac{\partial^2}{\partial x_\sigma^2} + \dfrac{1}{2}m_\sigma\omega_\sigma^2x_\sigma^2$. The intra-species and inter-species two-body interaction integrals are defined respectively as follows
\begin{align}
    U_{ijk\ell}^{B} = g_{B} \iint &\psi^*_{B,i}(x_{B,1})\psi^*_{B,j}(x_{B,2}) \delta(x_{B,1}-x_{B,2}) \cdot \nonumber\\
    &\psi_{B,k}(x_{B,1})\psi_{B,\ell}(x_{B,2})   dx_{B,1} dx_{B,2},\\
    U_{ijk\ell}^{BC} = g_{BC} \iint &\psi^*_{B,i}(x_B)\psi^*_{C,j}(x_C) \delta(x_B-x_C) \cdot \nonumber\\
    &\psi_{B,k}(x_B)\psi_{C,\ell}(x_C)   dx_B dx_C.
\end{align}
The many-body wave function for a two-component system can be expressed as
\begin{equation} \label{many-body wavefunction ansatz}
    |\Psi\rangle = \sum_{i_B=1}^{D_B}\sum_{i_C=1}^{D_C} c_{i_B,i_C}|n_B\rangle_{i_B}|n_C\rangle_{i_C},
\end{equation}
where $c_{i_B,i_C}$ are the expansion coefficients, $|n_\sigma\rangle_{i_\sigma}$ represent the Fock configurations, and $D_\sigma$ denotes the total number of configurations for component $\sigma$. The Fock basis for $N_\sigma$ particles distributed across $M_\sigma$ energy modes is defined as
\begin{equation}
    |n_\sigma\rangle = |n_{\sigma,1},n_{\sigma,2},\ldots,n_{\sigma,j},\ldots,n_{\sigma,M_\sigma} \rangle,
\end{equation}
where $n_{\sigma,j}$ is the number of particles from component $\sigma$ occupying the $j$-th single-particle state $\psi_{\sigma,j}(x_\sigma)$. By numerically diagonalizing the matrix representation, one can find the initial state of the battery set up $|\Psi(0)\rangle$ as the ground-state of the many-body Hamiltonian \eqref{eq:H0-2ndquantized}, and its quench dynamics can be computed as 
\begin{equation}
    |\Psi(0)\rangle = \sum_i \langle \Phi_i|\Psi(0)\rangle\exp\left(-\dfrac{iE_it}{\hbar} \right) |\Phi_i\rangle,
\end{equation}
where $|\Phi_i\rangle$ and $E_i$ are eigenpairs of \eqref{eq:H1-2ndquantized}. Because of the spatial symmetry of the single-particle wave functions $\psi_n(x)$, the initial state $|\Psi(0)\rangle$ is only spanned in the odd-parity Fock subspace. Therefore, one can significantly reduce the dimension of the Hilbert space by only considering odd-parity Fock states to build the Hamiltonian \cite{10.21468/SciPostPhys.15.2.048}.

\bibliography{apssamp}

\end{document}